\documentclass[twocolumn]{aastex631}

\begin{document}

\title{Discovery of a Rare Eclipsing Be/X-ray Binary System, Swift J010902.6-723710 = SXP 182}

\author[0009-0000-6957-8466]{Thomas M. Gaudin}
\affiliation{Pennsylvania State University}

\author[0000-0002-6745-4790]{Jamie A. Kennea}
\affiliation{Pennsylvania State University}

\author[0000-0002-0763-8547]{M.J. Coe}
\affiliation{Physics and Astronomy, The University of Southampton, SO17 1BJ, UK}

\author[0000-0002-4754-3526]{I. M. Monageng}
\affiliation{South African Astronomical Observatory, PO Box 9, Observatory, Cape Town 7935, South Africa}
\affiliation{Department of Astronomy, University of Cape Town, Private Bag X3, Rondebosch 7701, South Africa}

\author[0000-0001-5207-5619]{Andrzej Udalski}
\affiliation{Astronomical Observatory, University of Warsaw, Al. Ujazdowskie 4, 00-478 Warszawa, Poland}

\author[0000-0001-9788-3345]{L. J. Townsend}
\affiliation{Southern African Large Telescope, PO Box 9, Observatory, Cape Town 7935, South Africa}
\affiliation{South African Astronomical Observatory, PO Box 9, Observatory, Cape Town 7935, South Africa}

\author{David A.H. Buckley}
\affiliation{South African Astronomical Observatory, PO Box 9, Observatory, Cape Town 7935, South Africa}

\affiliation{Southern African Large Telescope, PO Box 9, Observatory, Cape Town 7935, South Africa}

\affiliation{Department of Astronomy, University of Cape Town, Private Bag X3, Rondebosch 7701, South Africa}

\affiliation{Department of Astronomy, University of the Free State, PO
Box 339, Bloemfontein, Cape Town, 9300, South Africa}

\author{Phil A. Evans}
\affiliation{Leicester University, UK}



\begin{abstract}

We report on the discovery of Swift J010902.6-723710, a rare eclipsing Be/X-ray Binary system by the Swift SMC Survey (S-CUBED). Swift J010902.6-723710 was discovered via weekly S-CUBED monitoring observations when it was observed to enter a state of X-ray outburst on 10 October 2023. X-ray emission was found to be modulated by a 182s period. Optical spectroscopy is used to confirm the presence of a highly-inclined circumstellar disk surrounding a B0-0.5Ve optical companion. Historical UV and IR photometry are then used to identify strong eclipse-like features re-occurring in both light curves with a 60.623 day period, which is adopted as the orbital period of the system. Eclipsing behavior is found to be the result of a large accretion disk surrounding the neutron star. Eclipses are produced when the disk passes in front of the OBe companion, blocking light from both the stellar surface and circumstellar disk. This is only the third Be/X-ray Binary to have confirmed eclipses. We note that this rare behavior provides an important opportunity to constrain the physical parameters of a Be/X-ray Binary with greater accuracy than is possible in non-eclipsing systems.

\end{abstract}



\section{Introduction} \label{sec:intro}

Be/X-ray Binaries (BeXRBs) are a type of interacting High Mass X-ray Binary (HMXB) that contain a main sequence OBe star and a compact object, typically a neutron star (NS). These binaries are characterized by moderately eccentric elliptical orbits ($e \sim 0.3$), orbital periods on the order of $\sim$10-1000 days, and the presence of strong emission lines such as the Balmer series H$\alpha$ line in the optical spectrum of the OBe star (for a comprehensive review of BeXRBs, see \citet{2011Reig}). The optical emission lines are interpreted to be a strong indication of a geometrically thin circumstellar disk \citep{2019Rivinius} that surrounds the donor star and is variable in size. Interactions between the NS and disk are capable of producing intermittent X-ray outbursts, which are perhaps the most prominent feature of BeXRB systems. 

There are two types of outburst that can be produced by NS interactions with the circumstellar disk \citep{1986Stella, 2001Okazaki, 2011Reig}. Type I outbursts are periodic in nature and associated with NS-disk interactions that occur at the periastron passage in each orbit\citep{1986Stella, 2001Okazaki}. Type II outbursts are much more luminous than Type I outbursts \citep{2001Okazaki, 2011Reig} and can even reach super-Eddington luminosities, such as during the SMC X-3 outburst described by \citet{2017Townsend} and the SMC X-2 outburst that occurred in 2015 \citep{2023Jaisawal, 2022Roy}. These luminous outbursts are independent of the orbital phase of the NS at the time of outburst and can last for multiple orbits \citep{2011Reig}. Due to both the long duration and phase independence of these events, Type II outbursts are thought to be related to the growth and shape of the Be star's disk \citep{2014Martin, 2017Monageng} causing it to interact with the NS for a longer time.

Since X-ray outbursts in BeXRBs are dependent on the disk of their donor star, these systems are prone to experiencing long quiescent states during which they are hard to detect and identify \citep{2023Coe}. Swift J010902.6-723710 is an example of a system that escaped identification due to a long period of quiescence. This newly-discovered system resides in the Small Magellanic Cloud (SMC), a satellite galaxy of the Milky Way that is well-documented to have a large population of HMXBs \citep{2010Antoniou} due to an unusually-active period of star formation that occurred in the SMC's recent past \citep{2004Harris, 2014Rezaeikh}. New SMC BeXRBs are still occasionally discovered \citep{2023Maitra}, indicating that a hidden population of previously-undetected systems still exists within the dwarf galaxy.

In an effort identify X-ray outbursts and detect new sources, the Swift SMC Survey (S-CUBED) \citep{2018Kennea} has operated since 2016 to provide regular X-ray and UV monitoring of the SMC. This survey utilizes the Neil Gehrels Swift Observatory \citep{2004Gehrels} to tile the galaxy on a weekly cadence with 142 overlapping tiles and $\sim$1 minute spent on each tile. Data recorded by S-CUBED \textbf{are} automatically analyzed to flag new X-ray outburst events that are detected by \textit{Swift}'s X-ray Telescope (XRT; \citealt{2005Burrows}). S-CUBED also observes all tiles using the Ultraviolet/Optical Telescope (UVOT; \citealt{2005Roming}) in the \textit{uvw1} band centered at 2600\AA. The archive of X-ray and UV photometric data taken by S-CUBED has proved to be an invaluable resource for discovering previously-undetected BeXRBs \citep{2020Monageng, 2020Kennea, 2021Coe, 2021Kennea, 2022Monageng}. It was through an X-ray selected search of the S-CUBED archive that Swift J010902.6-723710 was first identified as a candidate source in the summer of 2023, several months before it entered outburst during October 2023. 

In this paper, we report on the recent X-ray outburst that was experienced by Swift J010902.6-723710 which confirms its status as a newly-identified BeXRB system \citep{2023bcoe}. In the following sections, we present the results of observations made both before and during outburst using Swift, the Optical Gravitational Lensing Experiment (OGLE), and the Southern African Large Telescope (SALT). Additionally, all results are combined and analyzed in an effort to obtain information about the orbital and physical properties of both the NS and its mass donor OBe companion. 


\section{Observations} \label{sec:obs}

\subsection{Swift - XRT} \label{subsec:xrt obs}

\begin{figure}
    \includegraphics[scale=0.4]{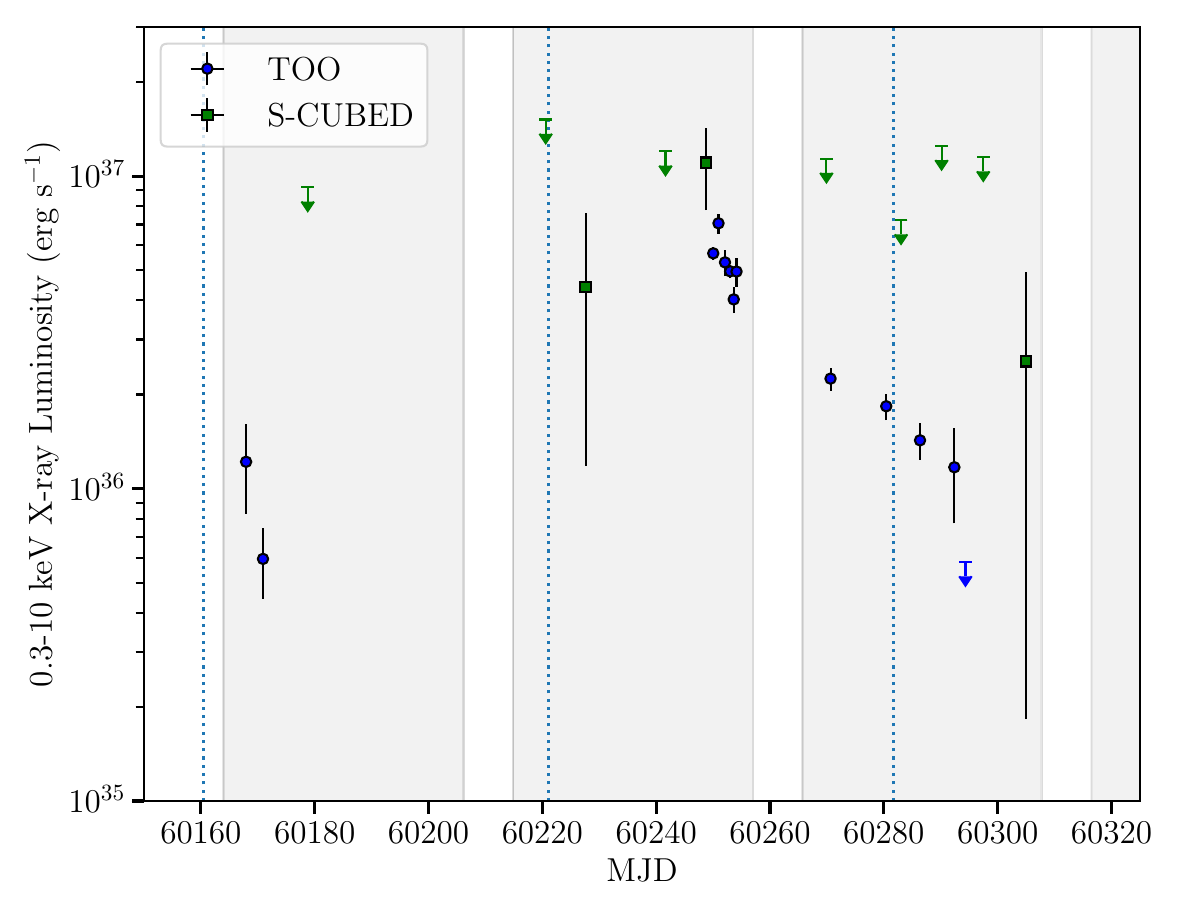}
    \caption{XRT light curve for Swift J010902.6-723710 from August 2023 to present, combining data from S-CUBED observations and follow-up TOOs. Note source visibility to Swift is indicated by the grey background, white vertical strips indicate periods when the target could not be observed. Vertical dotted lines mark predicted times for the periastron passages.}
    \label{fig:XRT lc}
\end{figure}

Swift J010902.6-723710 has remained in a quiescent state for most of the duration of S-CUBED monitoring.  XRT first detected this source during weekly S-CUBED monitoring in March of 2020. Two more detections occurred in July and September of 2021. This mini-flare of X-ray photons went un-reported due to its sparse detection frequency. The  X-ray luminosity of the flare peaked at an XRT count rate of $0.071_{-0.035}^{+0.051}$ counts s$^{-1}$ before returning to a quiescent state. The peak XRT count rate implies an approximate 0.3-10 keV band luminosity of $L_X = 4.76 \times 10^{36}$ erg s$^{-1}$ if the standard distance of 62.44 kpc \citep{2020Graczyk} to the SMC is assumed. 

The source was identified as a possible candidate BeXRB by examining the optical/IR SEDs of all unidentified X-ray sources in the S-CUBED survey (Gaudin et al. \textit{in prep.}). This identification motivated a deep 10ks Swift observation in August of 2023 to try and constrain the hardness ratio of the source's X-ray spectrum. However, the source was only weakly detected with a mean luminosity of $L_X = 9.07 \times 10^{35}$ erg s$^{-1}$. 

The results of all XRT observations during the recent outburst are shown in Figure \ref{fig:XRT lc}. Weekly S-CUBED monitoring detected X-ray emission from the source as it entered outburst on October 10th and again at peak XRT count rate on October 31st. The peak XRT count rate of Swift J010902.6-723710 was $0.22_{-0.058}^{+0.071}$ counts s$^{-1}$, implying a peak 0.3-10 keV band luminosity of $L_X = 1.35 \times 10^{37}$ erg s$^{-1}$ at a distance of 62.44 kpc. S-CUBED detection of outbursting behavior triggered deeper follow-up Swift observations of 5 ks per day on November 2nd and 4th in an effort to both monitor the outburst and constrain the spin period of the NS. Deep observations show the XRT count rate start to decline exponentially as expected, reaching a mean XRT count rate value of 0.11 counts s$^{-1}$ during the early November observations. Additional observations taken on November 22nd and December 2nd confirm the continuing trend of exponential decline in X-ray brightness as the outburst fades. However, weekly S-CUBED detections persisted, with additional detections occurring on December 12th and December 26th. These two detections had 0.3-10 keV luminosities of $L_X = 3.55 \times 10^{36}$ erg s$^{-1}$ and $L_X = 3.98 \times 10^{36}$ erg s$^{-1}$, respectively, which do not fit the trend of steadily declining flux values after the peak luminosity was reached. 

\begin{figure}
    \centering
    \includegraphics[scale=0.45]{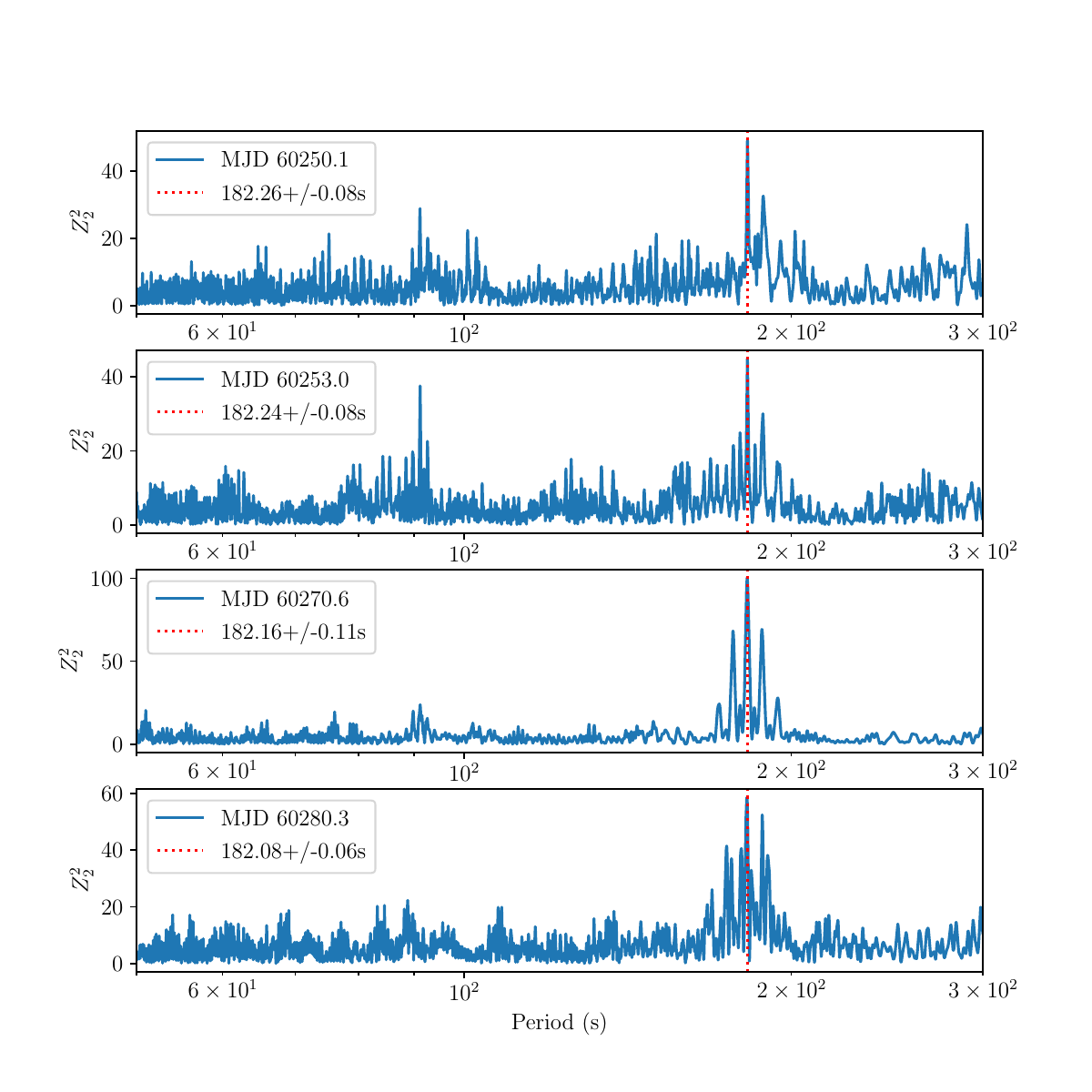}
    \caption{$Z_2^2$ Periodogram for four observations with exposures of $\simeq$5\,ks of Swift XRT data taken in PC mode. The periodogram shows a peak at 182 seconds in all four observations made after outburst was detected by S-CUBED weekly monitoring. There is evidence for pulsar spin-up over time, however the errors on period measurement are large meaning this result is not statistically significant.}
    \label{fig:x-ray periodogram}
\end{figure}

All 5 ks follow-up Swift observations were searched for the presence of periodic pulsations that are indicative of the spin period of the X-ray pulsar. The results of this search, carried out using a $Z_2^2$ Periodogram \citep{1983Buccheri}, are shown in Figure \ref{fig:x-ray periodogram}. In all 5 ks observations, the periodogram shows peaks at around $182$s (see Figure~\ref{fig:x-ray periodogram} for individual period measurements). Pulsar periods show small evidence for spin-up during the outburst, although given large errors on the periodicities due to the low numbers of counts and low (2.5s) PC mode frame time, the spin-up is not strongly detected. Folding the light curve of each individual XRT observation of 182s reveals an asymmetric double-peaked periodic signature, which is the expected shape for an X-ray pulsar light curve. Based on the phase-folded light curve and the strength of the periodogram peaks, we argue that the true spin period of the pulsar is 182s, allowing Swift J010902.6-723710 to be designated as SXP 182.

Spectral fitting is used to derive values for the spectral index of this source and column density along the line of sight. This is done by fitting the time-averaged 0.3-10 keV spectrum to an absorbed power-law model using the methods presented in \citet{2014Evans}. Using these methods, best-fit parameters and a 68\% confidence interval can be derived for both properties. SXP 182 is shown to have a hard X-ray spectrum, with a derived photon index for the source of $\Gamma = 0.52_{-0.15}^{+0.16}$, which is consistent with the photon index of $\Gamma  \sim 0-1$ expected of a BeXRB system \citep{2018Kennea}. Additionally, power law fitting indicates that the column density along the line of sight towards this source is $N_H = 5.1_{-1.9}^{+2.3} \times 10^{21}$ cm$^{-2}$ which is higher than the average value of the column density of $N_H = 5.34 \times 10^{20} \text{cm}^{-2}$ towards the SMC \citep{2013Willignale, 2018Kennea}. 

\begin{figure}
    \centering
    \includegraphics[scale=0.41]{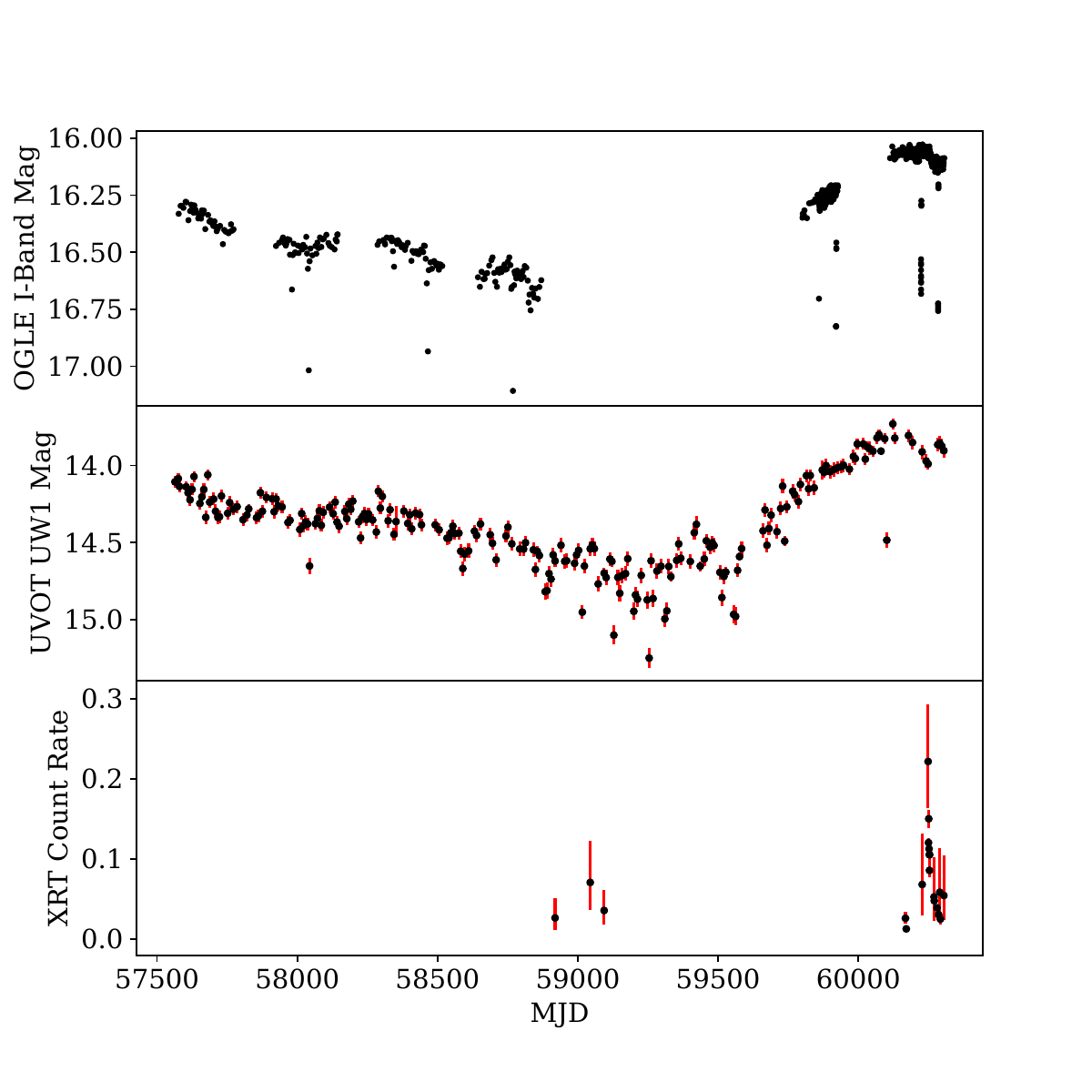}
    \caption{OGLE IV, Swift UVOT, and Swift XRT light curves for Swift J010902.6-723710.}
    \label{fig:all lcs}
\end{figure}

\subsection{Swift - UVOT} \label{subsubsec:uvot obs}

Swift J010902.6-723710 has been observed 221 times by UVOT in the \textit{uvw1}-band since 2016 as part of S-CUBED monitoring, which represents approximately weekly coverage. UVOT light curves are generated for S-CUBED sources using the FTOOLS software package \citep{1999Blackburn}. Photometric data are extracted from a circle with 5 arcsecond radius around the XRT source position for the object using the \texttt{uvotsource} method that is part of FTOOLS. Using this method, the light curve presented in Figure \ref{fig:all lcs} was generated for the entire duration of S-CUBED. 

UVOT data identifies Swift J010902.6-723710 as a persistent emitter in the \textit{uvw1} band as it is well-detected in every observation with a 14.4 mean magnitude. The most prominent feature of this light curve is the presence of strong UV variability on timescales of 500-1000 days. The source reached its minimum average magnitude of 14.7 in October 2021 after years of gradual dimming from its initial brightness of 14.1 at the start of S-CUBED monitoring. At this point, the source began a period of rapid brightening lasting for over 500 days, reaching a maximum magnitude of 13.7 on June 27, 2023. After the peak magnitude was reached, the source has entered a steep dimming phase that corresponds with the start of the XRT outburst. This dimming phase was observed to be concluded on MJD 60283, when the source was observed to be at its pre-outburst \textit{uvw1} magnitude.

The long period of UV brightening that Swift J010902.6-723710 underwent was crucial in identifying the system as a candidate BeXRB before the outburst occurred. This type of UV increase leading to outburst has been observed in other sources such as Swift J004516.6–734703 \citep{2020Kennea} which demonstrated very little UV variability before experiencing a similar brightening in the lead up to a Type I outburst. In the outburst of Swift J004516.6–734703, the UV brightening was interpreted to be the result of a circumstellar disk forming around the companion Be star in the newly-identified binary. For Swift J010902.6-723710, we can interpret the long-term variability to be an indication of growth and decay in the size of the disk, where the disk expanded over the last 500+ days to a large-enough radius for NS-disk interactions to occur. 

\subsection{OGLE - the Optical Gravitational Lensing Experiment} \label{subsec:ogle obs} 
The OGLE project \citep{Udalski2015} provides long term I-band photometry with a cadence of 1-3 days for sources in the Magellanic Clouds. From the X-ray position the optical/IR counterpart to Swift J010902.6-723710 is identified as 
2MASS J01090226-7237101. It was observed continuously for nearly 3 decades by OGLE, though the observations were interrupted for approximately 3 years during COVID-19. 

The counterpart to the X-ray source is identified in the OGLE catalogue as:\\
\\
OGLE II (I-band): smc\_sc11.107571 \\
OGLE III (I-band): smc110.3.22311 \\
OGLE IV (I band): smc726.26.15515 \\
OGLE IV (V band):  smc726.26.v.22358 \\

The I band data from just the OGLE IV project are shown in the top panel of Figure \ref{fig:all lcs} for comparison with the contemporaneous Swift UVOT and XRT data.

\begin{figure}
    \centering
    \includegraphics[scale=0.42]{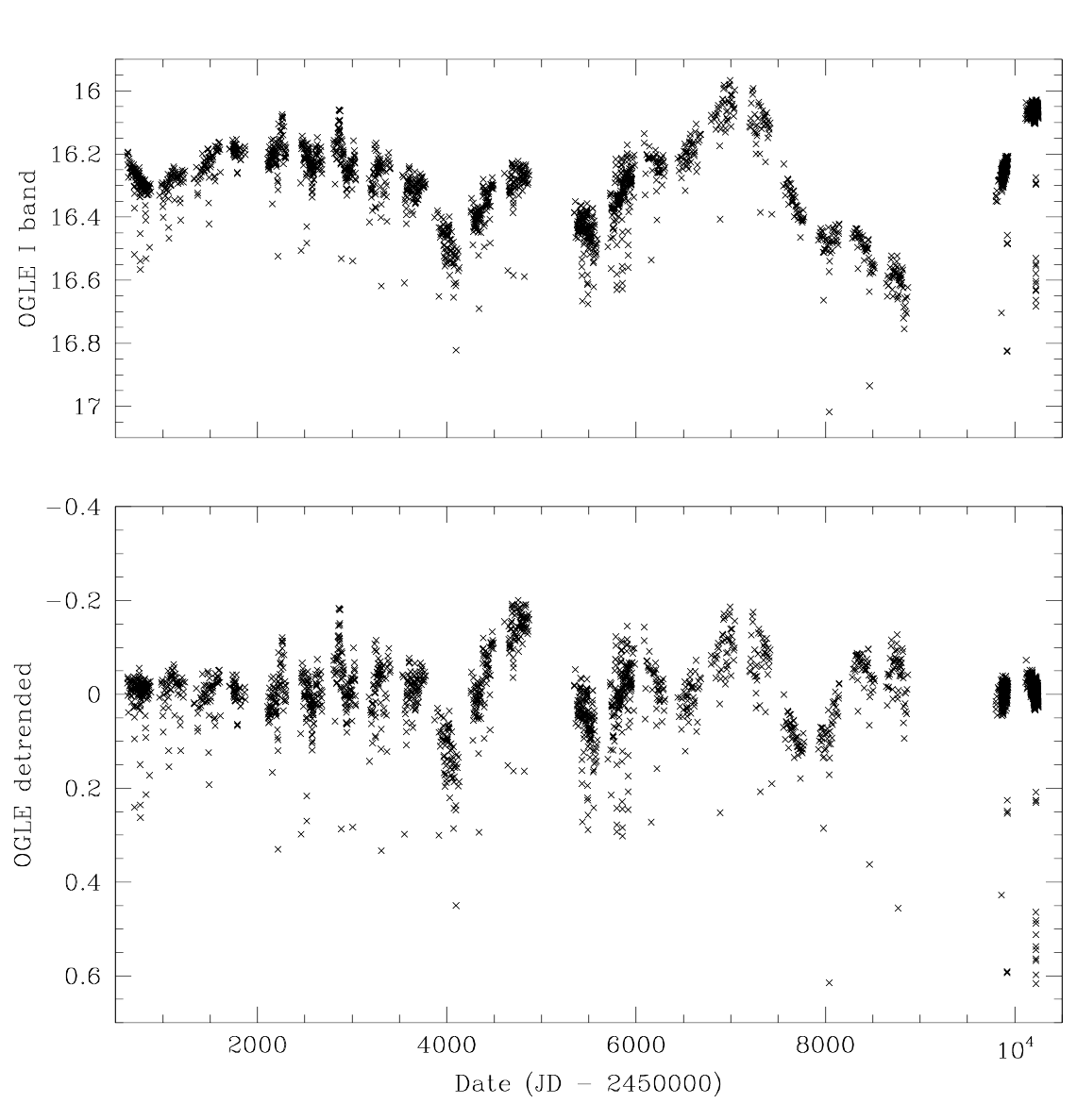}
    \caption{Top panel - the whole OGLE data set covering 26 years. Lower panel - the above data detrended with a simple polynomial in preparation for timing analysis.}
    \label{fig:allog}
\end{figure}

\begin{figure}
    \centering
    \includegraphics[scale=0.38]{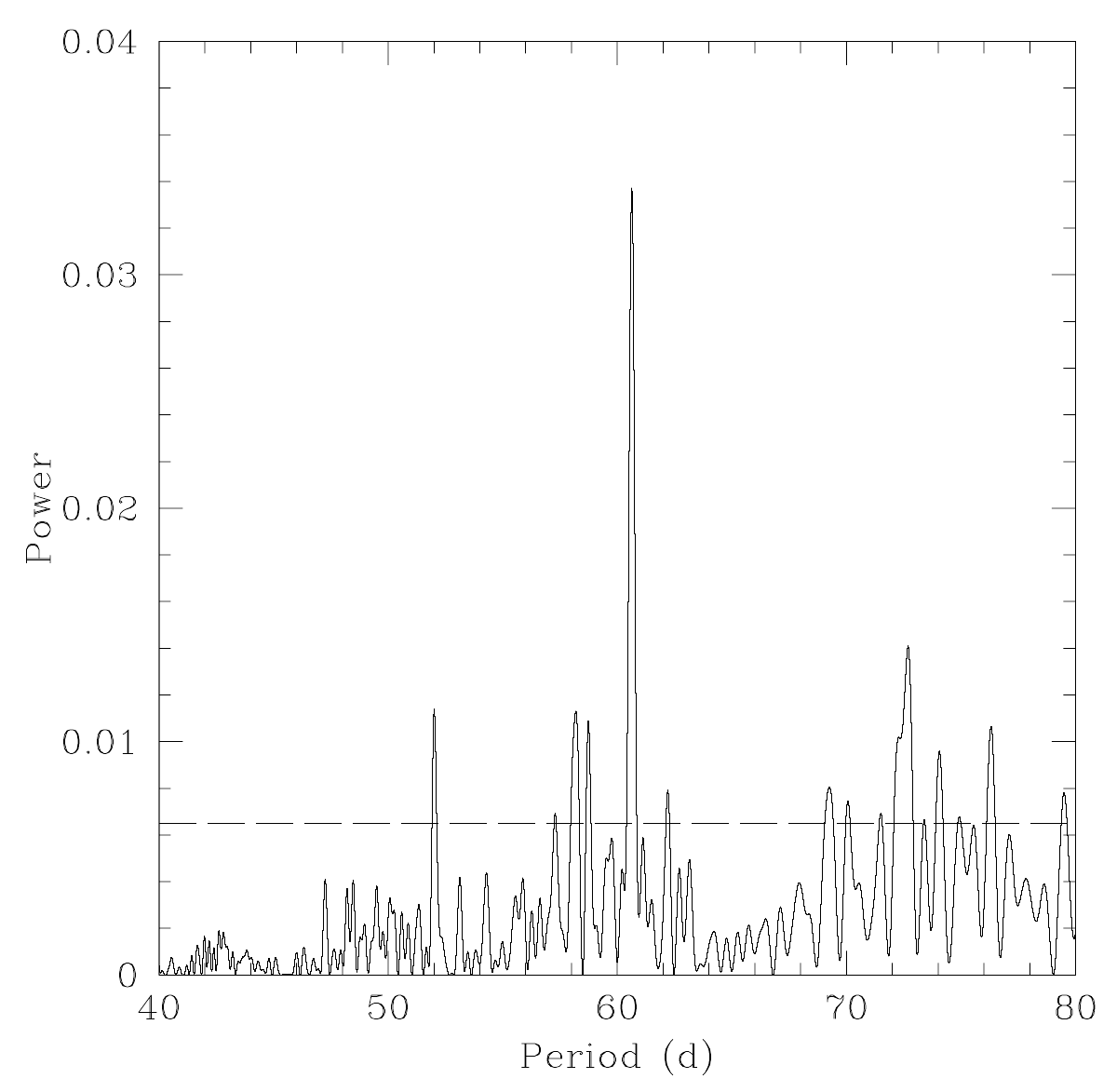}
    \caption{Generalised Lomb-Scargle analysis of the detrended OGLE data set. The peak is at a period of $60.623\pm0.001$ days.The horizontal dashed line represents the False Alarm Probabilty of 1\%.}
    \label{fig:ls4080}
\end{figure}


\begin{figure}
    \centering
    \includegraphics[scale=0.42]{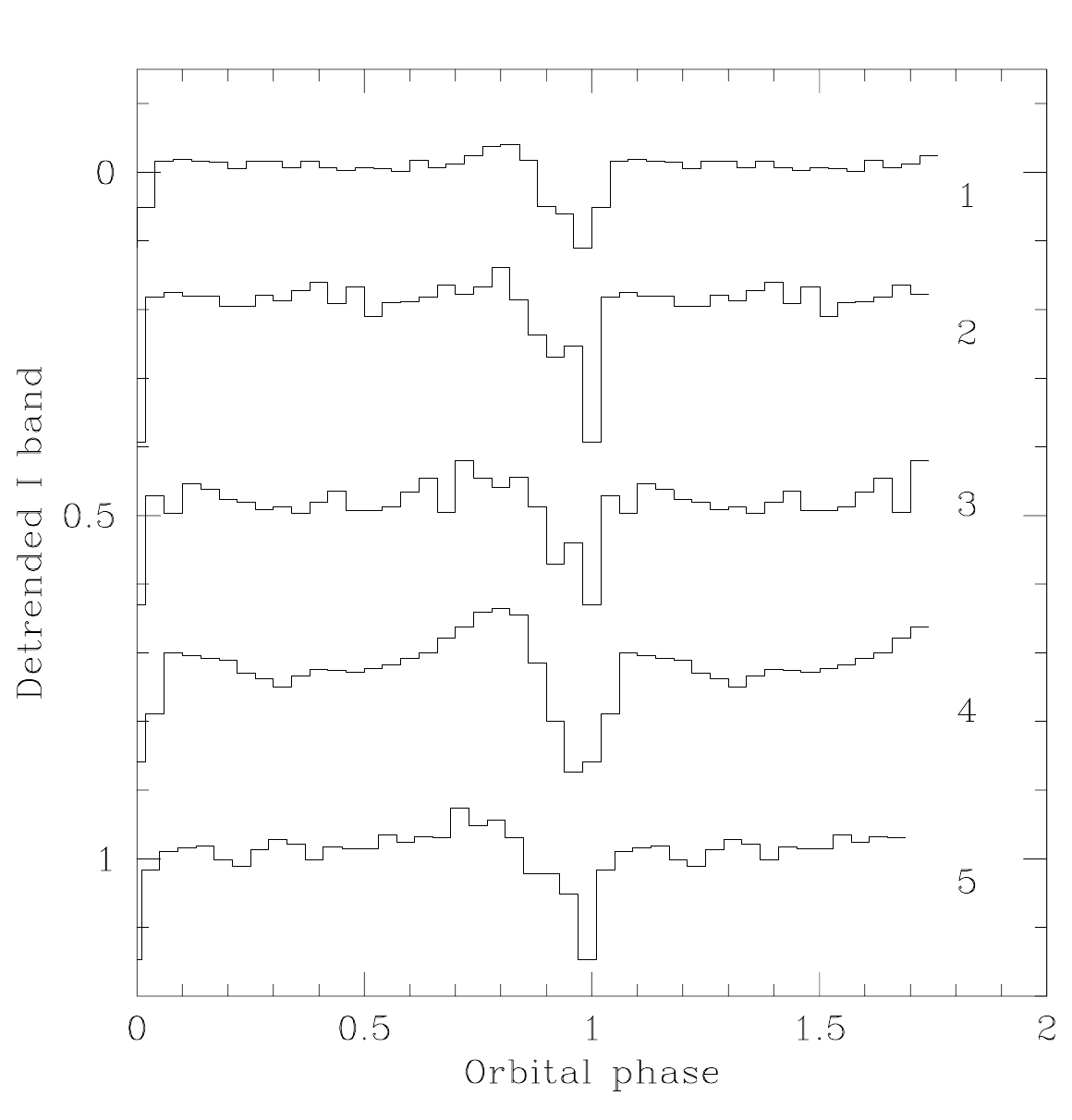}
    \caption{Detrended OGLE data divided into 5 consecutive epochs (see Table \ref{tab:ogle}) and each then folded using the ephemeris given in Equation \ref{eq:1}. For the purpose of being able to see the separate profiles, those numbered 2-4 have been arbitrarily shifted in the y-axis such as to avoid overlap.}
    \label{fig:fd25multi}
\end{figure}

\begin{table}
	\centering
	\caption{OGLE data date ranges used in Figure \ref{fig:fd25multi}}
    \setlength\tabcolsep{2pt}
	\begin{tabular}{ccc} 
		\hline\hline
		Data block & Dates & Average I band  \\
            number & JD-2450000& magnitude \\
		\hline
            1 & 600-2000& 16.26 \\
            2 & 2000-3100& 16.22 \\
            3 & 3100-5000& 16.35 \\
            4 & 5200-6400& 16.37\\
            5 & 6400-9000& 16.34\\
    
		\hline
	\end{tabular}
	\label{tab:ogle}
\end{table}

For the purpose of period analysis the whole of the 26.4 years of OGLE data were used and first detrended using a simple polynomial - see Figure \ref{fig:allog}. This data set was then analysed with a generalised Lomb-Scargle routine and the resulting power spectrum is shown in Figure \ref{fig:ls4080}. The peak in the power spectrum is at a period of $60.623\pm0.001$ days. This peak is driven by the sharp eclipse-like features that can be clearly seen by eye in the top panels of Figures \ref{fig:all lcs} and  \ref{fig:allog}. It is assumed that this represents the orbital period of the system.

If all the detrended OGLE data are folded at the proposed binary period, with a phase 0.0 set to the date of the first OGLE measurement (JD 2450627.9) then the sharp eclipse-like dips are clearly seen in Figure \ref{fig:fd25multi}. The average FWHM of the dips is 0.1 in phase. It is also clear that the ingress into the eclipse is much less sharp than the egress. To explore the changes in the shape of the eclipse over time the OGLE data were divided up into 5 separate time segments representing the times when visible changes were occurring in the profile. The resulting separate profiles are shown in Figure \ref{fig:fd25multi}. The 5 epochs chosen are listed in Table \ref{tab:ogle}. Using the phase of the dip in the I-band shown in Figure \ref{fig:fd25multi} as a reference point, the ephemeris for the time of the optical eclipses is given by Equation \ref{eq:1}: 

\begin{equation}
T_{ecl} = 2450645.1 + N(60.62\pm{0.01}) ~\textrm{~JD}
\label{eq:1}
\end{equation}

\begin{figure*}
    \centering
    \includegraphics[width=0.45\textwidth]{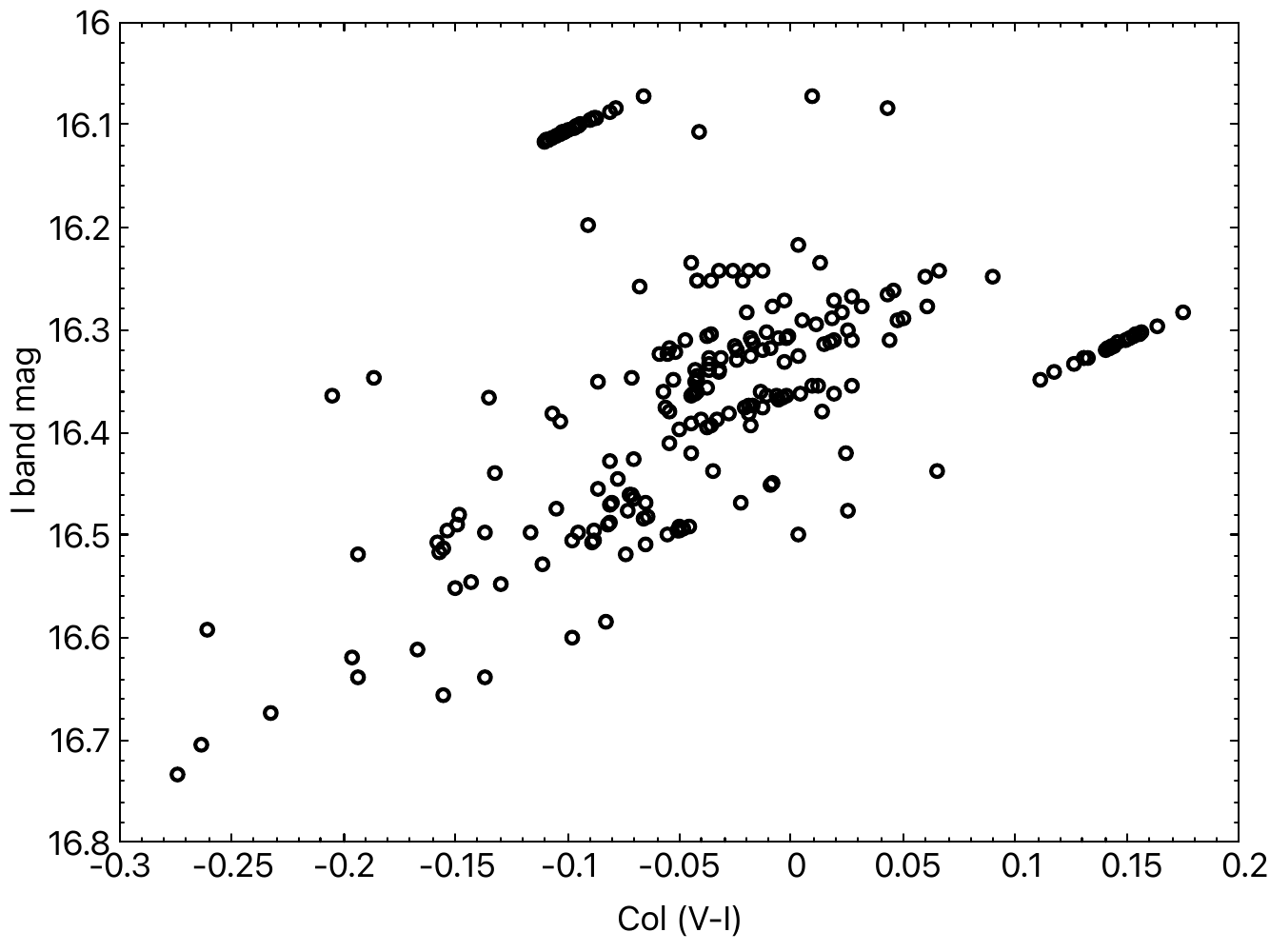}
    \hfill
    \includegraphics[width=0.45\textwidth]{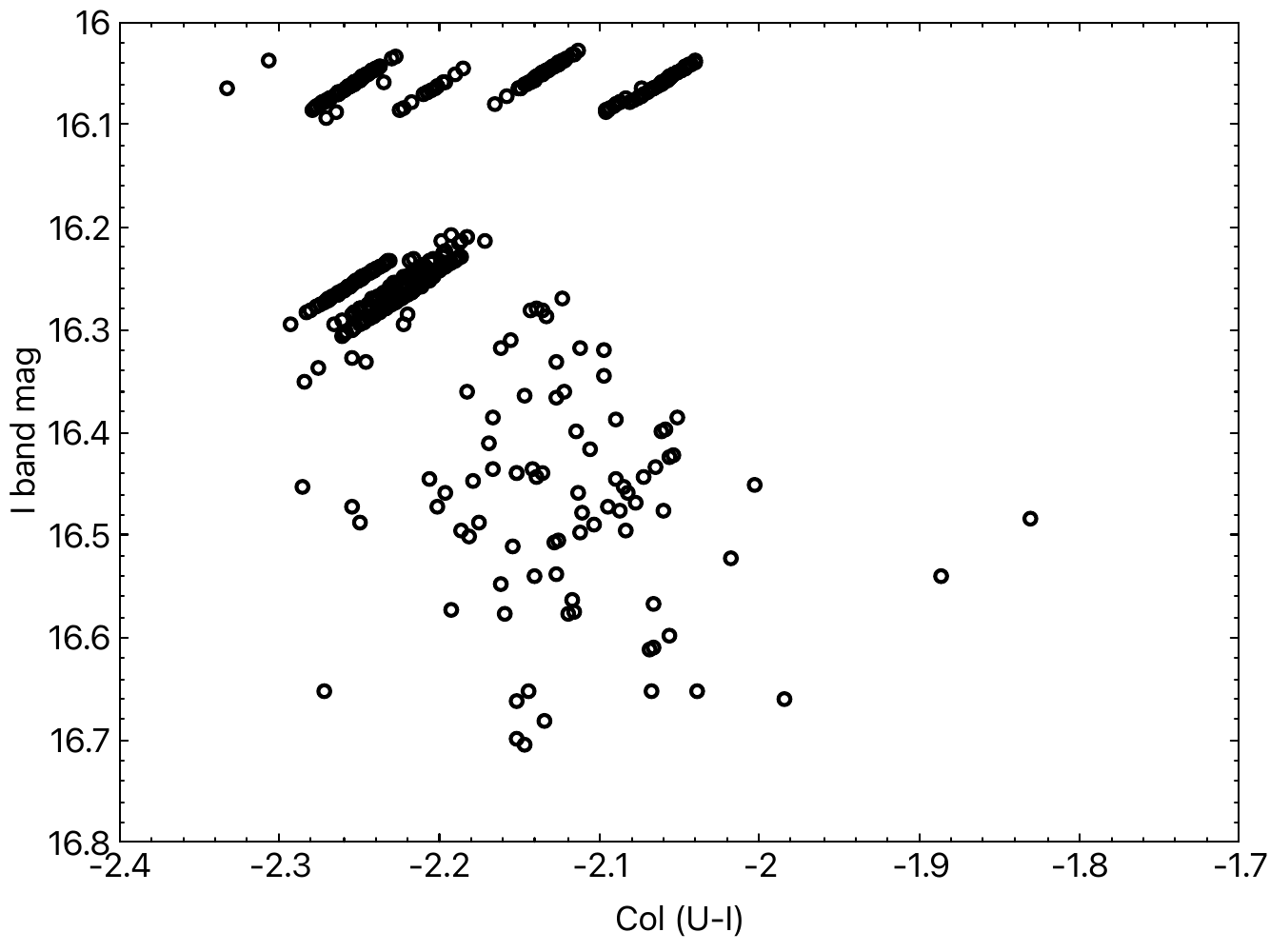}
    
    \caption{$(V-I) - I$ (left) and $(U-I) - I$ (right) colour-magnitude diagrams of Swift J010902.6-723710}
    \label{fig:CMDs}
\end{figure*}

In addition to the regular I-band measurements the OGLE IV project also records V band magnitudes every few days. This provides the opportunity to investigate the overall colour changes seen in the system as function of brightness - a colour-magnitude diagram (CMD). Since the V band measurements are less frequent than the I band, the determination of (V-I) can only occur when the I and V measurements are close enough together in time. In this instance the proximity of the 2 measurements is set to be less than 3 days. This means that occasionally one V band measurement may be partnered with more than one I band measurements. The result is shown in Figure \ref{fig:CMDs} and discussed below.

\subsection{SALT - the Southern African Large Telescope} \label{subsec:salt obs}

\label{subsec:inclination}

\begin{table}
	\centering
	\caption{The H$\alpha$ equivalent width (EW) and peak separation ($\Delta V$) measured from SALT observations.}
    \setlength\tabcolsep{2pt}
	\begin{tabular}{ccccc} 
		\hline\hline
		Date & MJD & EW (\AA) & Grating & $\Delta V$ (km/s)  \\
		\hline
            03-11-2023 & 60251.88& -7.4$\pm$0.3 & PG0900 & $-$ \\
            09-11-2023 & 60257.87& -7.3$\pm$0.5 & PG2300 & 373.6$\pm$0.4 \\
            02-12-2023 & 60280.85& -7.4$\pm$0.4 & PG2300 & 383.4$\pm$0.9 \\
		\hline
	\end{tabular}
          	\label{tab:EW}
\end{table}

Swift J010902.6-723710 was observed with the Southern African Large Telescope (SALT; \citealt{2006Buckley}) with the Robert Stobie Spectrograph \citep{2003Burgh,2003Kobulnicky} using different settings. On 03-11-2023 (MJD60251.88) the PG0900 grating was used (grating angle of 15.125$^{\circ}$) with an exposure time of 1200 sec covering a wavelength range of $4200-7250$~\AA. The PG2300 grating was used on 09-11-2023 (MJD60257.87) and 02-12-2023 (MJD60280.85) (grating angle of 48.875$^{\circ}$) with an exposure time of 1200 sec covering a wavelength range $6100-6900$~\AA. An additional observation was taken on 09-11-2023 (MJD60257.87) with the PG2300 grating (grating angle of 30.5$^{\circ}$) with an exposure time of 1500 sec covering a wavelength range $3840-4915$~\AA. The primary reductions, which include overscan corrections, bias subtraction, gain and amplifier cross-talk corrections, were performed with the SALT science pipeline \citep{2012Crawford}. The remainder of the data reduction steps, which comprise wavelength calibration, background subtraction, and extraction of the one-dimensional spectrum, were done in \textsc{iraf}. All spectra were corrected for the heliocenter and redshift of the SMC of 145.6~km/s \citep{McConnachie2012}.

Figure~\ref{fig:Ha_profiles} shows the H$\alpha$ emission line profiles. The observations obtained on MJD60257.87 and MJD60280.85 with the PG2300 grating exhibit asymmetric double-peak profiles as a result of the Keplerian distribution of matter in the disc when the disc is viewed at non-zero inclination angles. The observation performed on MJD60251.88 with the PG0900 grating shows an asymmetric single-peak profile since the resolution of the grating is insufficient to resolve the two peaks. The H$\alpha$ equivalent width measurements are recorded in Table~\ref{tab:EW}.

\section{Discussion} \label{sec:discussion}

\subsection{Corbet Diagram} \label{subsec:corbet}

\begin{figure}
    \centering
    \includegraphics[scale=0.41]{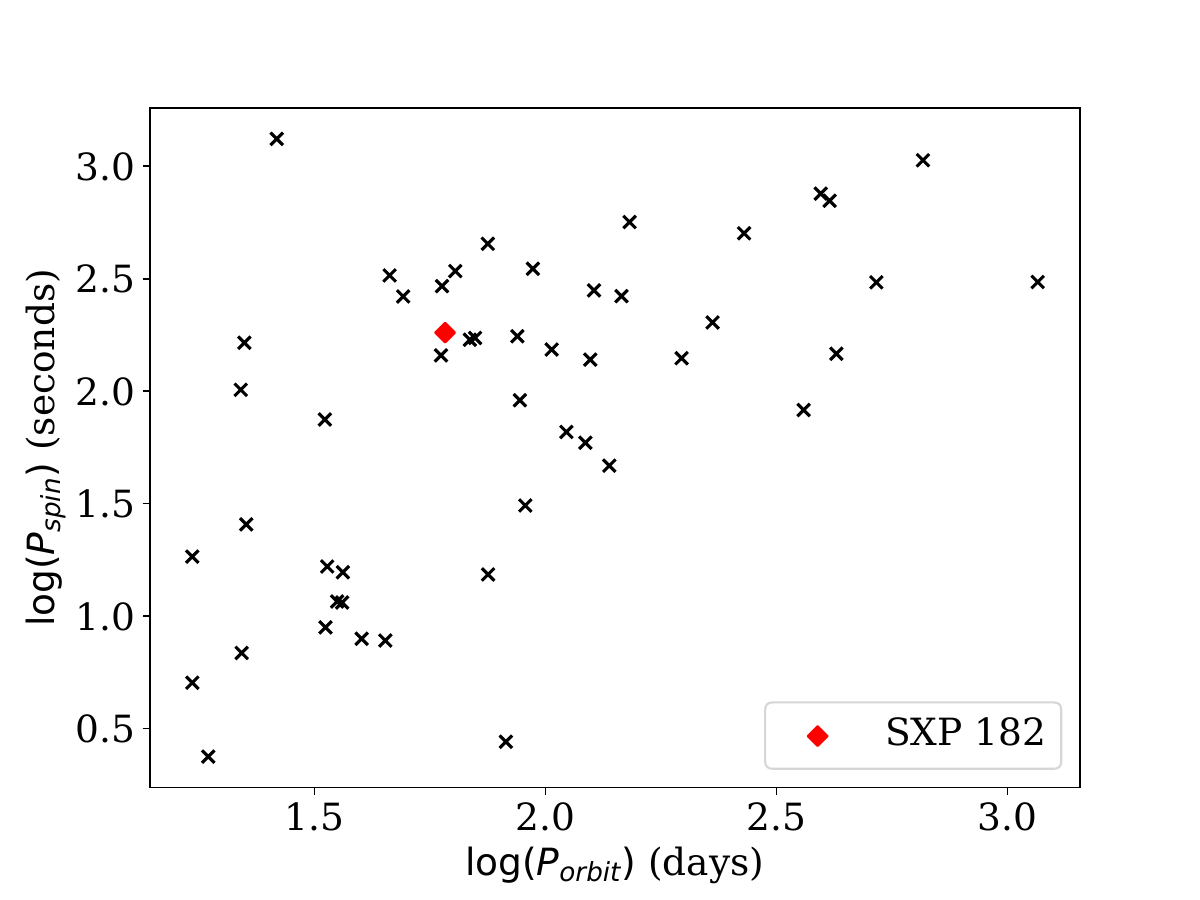}
    \caption{Corbet diagram for all BeXRBs in the SMC with known orbital periods and spin periods \citep{2005Haberl, 2015coekirk, 2017McBride, 2017Carpano, 2018Kennea, 2019Lazzarini, 2020Kennea, 2022Carpano, 2023Maitra}. SXP 182 is plotted alongside these data and is shown to demonstrate the same correlation between X-ray pulsar spin period and NS orbital period that is seen in the rest of the SMC BeXRB population.}
    \label{fig:corbet}
\end{figure}

Combining the spin period with the orbital period allows this new system to be placed on the Corbet diagram. The original Corbet diagram \citep{1984Corbet} showed a correlation between the spin and orbital periods of neutron stars in BeXRBs. However, instead of a correlation between spin and orbital periods, a modern version (shown in Figure \ref{fig:corbet}) shows that these properties are well-constrained to a specific region of the Corbet diagram. BeXRB systems are expected to fall above the diagonal of the diagram, which represents a log-linear relationship between the orbital period in days and the spin period in seconds of the NS. Figure \ref{fig:corbet} shows the location of SXP 182 when placed on the Corbet diagram with all BeXRBs in the SMC that have known spin and orbital periods \citep{2005Haberl, 2015coekirk, 2016Haberl, 2017McBride, 2017Carpano, 2018Kennea, 2019Lazzarini, 2020Kennea, 2022Carpano, 2023Maitra}. The location of SXP 182 is consistent with the trend shown by other SMC BeXRBs. This serves as a check that the periods derived in Sections \ref{subsec:xrt obs} and \ref{subsec:ogle obs} are consistent with the observational properties of other BeXRBs. For Swift J010209.6-723710, the derived orbital period of 60.623 days and pulsar spin period of 182 seconds place the system near the center of the distribution of orbit and spin periods for SMC BeXRBs, providing a strong piece of evidence validating the source as a newly-discovered BeXRB.

\subsection{Inclination angle of the Be disc}

\begin{figure}
    \centering
    \includegraphics[scale=0.37]{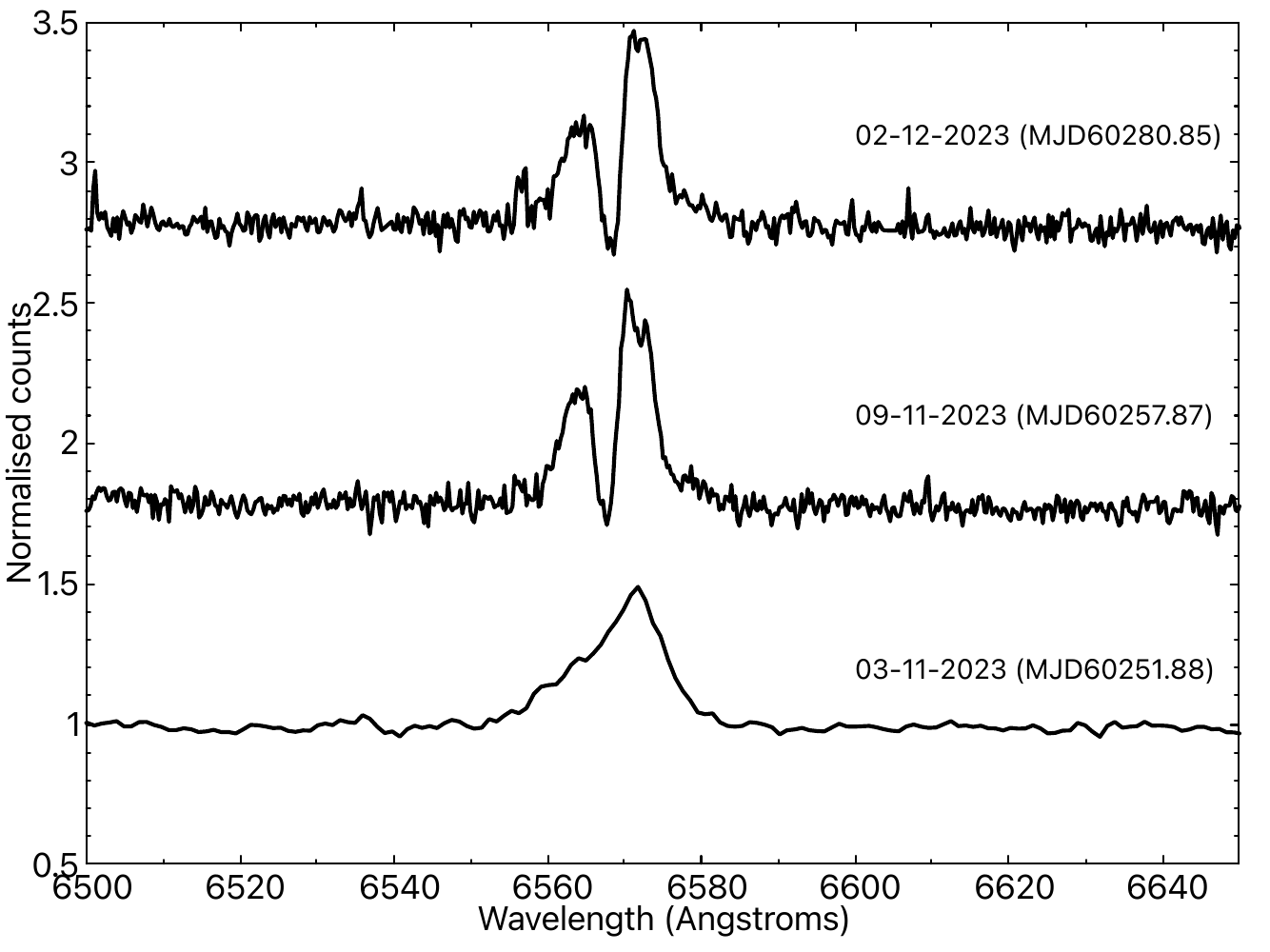}
    \caption{The evolution of the H$\alpha$ emission line from SALT observations. The spectra are corrected for the heliocenter and redshift of the SMC.}
    \label{fig:Ha_profiles}
\end{figure}

The peak separation of the double-peak H$\alpha$ emission lines (Figure~\ref{fig:Ha_profiles}) can be used to estimate the size of the H$\alpha$ emitting region \citep{Huang1972}:
\begin{equation}
    R = \frac{GM_\ast \sin^2 i}{(0.5\Delta V)^2}
\end{equation}
Using $M_\ast = 17.8$~M$_\odot$ \citep{Cox2000} based on the spectral type of the massive companion, this results in a disc size radius range of $92 - 97 \sin^2 i$~R$_\odot$. We can estimate the inclination angle of the Be disc, assuming that the disc and orbital planes are aligned and that during the period of X-ray activity the NS was accreting matter from the outermost parts of the Be disc at periastron passage. The semi-major axis of the orbit can be estimated by assuming a companion mass of $M_\ast = 17.8$~M$_\odot$ and a neutron star mass of $M_{NS} = 1.4$~M$_\odot$. Similarly, the periastron passage can be estimated using a conservative value for the eccentricity of $e\sim0.5$ based on the relationship between eccentricity and orbital period presented in \citet{2011Townsend}. This results in a range of disc inclination angles of $72 - 90^\circ$. The suggestion of a high disc inclination angle is corroborated by the H$\alpha$ emission line displaying a double peak morphology with a deep central depression from the high-resolution observations. 

Figure~\ref{fig:CMDs} shows the $(V-I) - I$ and $(U-I) - I$ colour-magnitude diagrams. The general trend of the colour-magnitude plots is a redder-when-brighter pattern, which is indicative of inclination angles below $90^\circ$\citep{Harmanec1983, Rajoelimanana2011, Reig2015}. This trend is more noticeable in the $(V-I) - I$ plot where the range in colour is broader since the simultaneous $V$ and $I$ band observations were taken during a period of substantial optical variability. 

\subsection{Spectral Classification  of the massive counterpart}
\begin{figure*}
    \centering
    \includegraphics[scale=0.65]{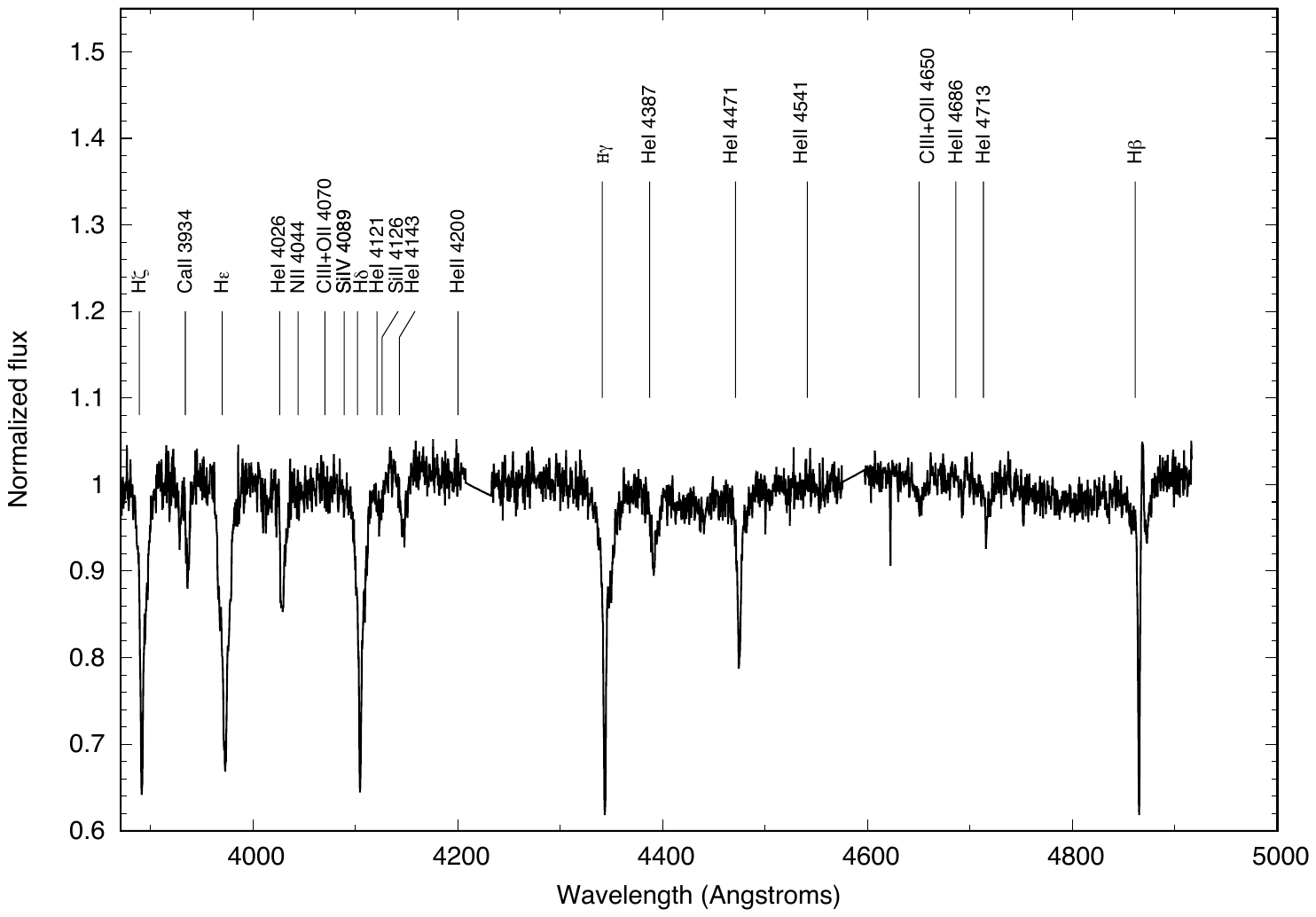}
    \caption{The SALT spectrum of Swift J010902.6-723710 covering the blue region with different line species labeled at their expected rest wavelengths. The spectrum is corrected for the heliocenter and redshift of the SMC.}
    \label{fig:SALT_blue}
\end{figure*}

Figure~\ref{fig:SALT_blue} shows the SALT spectrum of Swift J010902.6-723710 covering the blue wavelength range. The spectrum shows clear signatures of an early-type star with several Balmer and helium lines present. The H$\beta$ line is in weak emission with an asymmetric profile, along with the other absorption lines exhibiting infilling due to the presence of the circumstellar disc. According to the criteria presented in \cite{Evans2004}, the strong presence of the He I lines at 4026, 4143, 4388 and 4471~\AA~ as well as the weak presence of the He II lines at 4541 and 4686~\AA~ constrains the spectral type to B0-0.5. The faintest V-band observation from OGLE measurements during our period of monitoring is $\sim$16.5 magnitudes. Using this and the distance modulus of the SMC of 18.95 \citep{Graczyk2013}, this results in a luminosity class of V \citep{Straizys1981, Pecaut2013}. In summary, the spectral class of the massive companion in Swift J010902.6-723710 is B0-0.5 V. 

\label{subsec:spectral class}

\subsection{Eclipsing Behavior} \label{subsec:eclipses}

\begin{figure}
    \centering
    \includegraphics[scale=0.59]{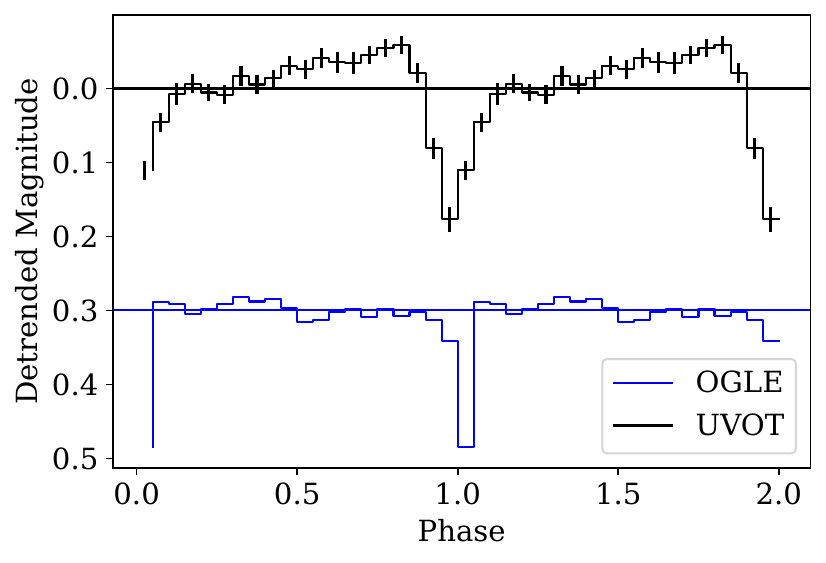}
    \caption{A comparison of the OGLE IV and UVOT light curves folded at the proposed binary period of 60.623 days using Equation \ref{eq:1}. The OGLE data are offset vertically to avoid overlap.}
    \label{fig:uv ir eclipse}
\end{figure}

The orbital period of Swift J010902.6-723710 derived from long-term OGLE monitoring can be used to re-examine the UVOT light curve in search of periodic behavior. This was done by de-trending the UVOT data using a 5th-order polynomial and folding the de-trended data at the 60.623 day orbital period of the system. Figure \ref{fig:uv ir eclipse} shows the comparison between the binned and time-averaged OGLE and UVOT light curves starting at MJD 57563, which is the date of the first S-CUBED observation. Despite S-CUBED's weekly cadence providing infrequent sampling of any singular orbit, the eclipsing behavior is clearly visible in the de-trended and folded UVOT light curve. When compared to the OGLE data, it becomes evident that the shape of the eclipse profile is wavelength-dependent. The UVOT data shows increasing emission that peaks just before each eclipse begins, which is similar to the behavior exhibited by the system during the 4th OGLE data block shown in Figure \ref{fig:fd25multi}. Over the lifetime of S-CUBED, the UVOT eclipse profile is also shown to be broader than the OGLE eclipse profile with similar depth at the time of maximum eclipse. At the maximum depth of the eclipse, the \textit{uvw1}-band magnitude drops by an average of 0.22 magnitudes and I-band magnitude decreases by an average of 0.25 magnitudes. Converting these magnitude decreases to flux measurements for both telescopes, the implied relative decrease in flux is at both wavelengths $\frac{\Delta F}{F} \sim 0.2$. 

Equation 2 from \citet{2013Maggi} can be used to estimate the size of the accretion disk based on the relative decrease in flux that is observed during eclipse:
\begin{equation}
    \frac{\Delta F}{F} = \left( \frac{R_X}{R_C} \right)^2
\end{equation}
where $R_X$ is the radius of the eclipsing object and $R_C$ is the radius of the optical companion. The value for $R_X$ can be estimated due to the constraints placed on $R_C$ via the results of Section \ref{subsec:spectral class}. The typical radius for a spectral class B0V star is 7.4 $R_\odot$ \citep{1976allen}. This value can be used to constrain the upper limit for the radius of the eclipsing object at $R_X = R_C\sqrt{\frac{\Delta F}{F}}  = 3.3 \, R_\odot$. A radius of this size rules out the NS, a large planet, or a Sun-like tertiary star as the cause of the eclipsing behavior. Therefore, it can be concluded that the eclipsing object is likely an extended accretion disk that surrounds the NS. \citet{2013Maggi} derives the theoretical upper limit for the size of an accretion disk in a BeXRB system to be $r_c \sim 11.5 \, R_\odot$, assuming Bondi-Hoyle accretion and a 10 $M_\odot$ companion star. The 3.3 $R_\odot$ accretion disk size derived for Swift J010902.6-723710 is well within this upper limit value, providing further evidence in favor of an eclipsing NS accretion disk.

Very few BeXRBs have been observed to demonstrate eclipsing behavior. Swift J010902.6-723710 is only the third known eclipsing BeXRB system and the second to be found in the SMC. LXP 168.8 \citep{2013Maggi} was found to have a eclipsing accretion disk with a 24.3 day orbital period. SXP 5.05 \citep{2015Coe2} was observed to contain a Be star that eclipses its NS companion every 17 days. This newly-discovered eclipsing system thus provides a unique opportunity to further constrain the physical parameters of the system such as the sizes of both disks and the masses of both stars. More observations are needed to further characterize the BeXRB using subsequent eclipses.

\subsection{Outburst Type} \label{subsec:outburst}

\renewcommand{\arraystretch}{1.25}
\begin{table}
    \centering
    \begin{tabular}{cccc}
    \hline\hline
        Date & MJD & Phase & $L_{X}$ (erg s$^{-1}$) \\
        \hline
        \textbf{10-10-2023} & 60227 & 0.06 & $4.14_{-2.61}^{+4.27} \times 10^{36}$ \\
        10-31-2023 & 60248 & 0.40 & $1.35_{-0.391}^{+0.475} \times 10^{37}$ \\
        11-02-2023 & 60250.0 & 0.47 & $5.66_{-0.275}^{+0.275} \times 10^{36}$ \\
        11-02-2023 & 60250.9 & 0.48 & $7.06_{-0.527}^{+0.527} \times 10^{36}$ \\
        11-04-2023 & 60252.1 & 0.50 & $5.30 _{-0.508}^{+0.508} \times 10^{36}$ \\
        11-05-2023 & 60253.0 & 0.52 & $4.96_{-0.253}^{+0.253} \times 10^{36}$ \\
        11-05-2023 & 60253.6 & 0.53 & $4.03_{-0.385}^{+0.385} \times 10^{36}$ \\
        11-06-2023 & 60254 & 0.54 & $4.95_{-0.518}^{+0.518} \times 10^{36}$ \\
        11-21-2023 & 60269.9 & 0.80 & $3.20_{-2.04}^{+3.33} \times 10^{36}$ \\
        11-22-2023 & 60270.6 & 0.81 & $2.25_{-0.190}^{+0.190} \times 10^{36}$ \\
        12-02-2023 & 60280 & 0.96 & $1.83_{-0.177}^{+0.177} \times 10^{36}$ \\
        12-08-2023 & 60286 & 0.06 & $1.43_{-0.192}^{+0.192} \times 10^{36}$ \\
        12-12-2023 & 60290 & 0.13 & $3.55_{-2.28}^{+3.78} \times 10^{36}$ \\
        12-14-2023 & 60292 & 0.16 & $1.17_{-0.397}^{+0.397} \times 10^{36}$  \\
        12-26-2023 & 60304 & 0.36 & $3.98_{-2.26}^{+3.69} \times 10^{36}$ \\
        \hline
    \end{tabular}
    \caption{S-CUBED and Swift TOO XRT detection dates, orbital phases from Equation \ref{eq:1}, and luminosities during outburst. }
    \label{tab:XRT data}
\end{table}

\begin{figure}
    \centering
    \includegraphics[scale=0.45]{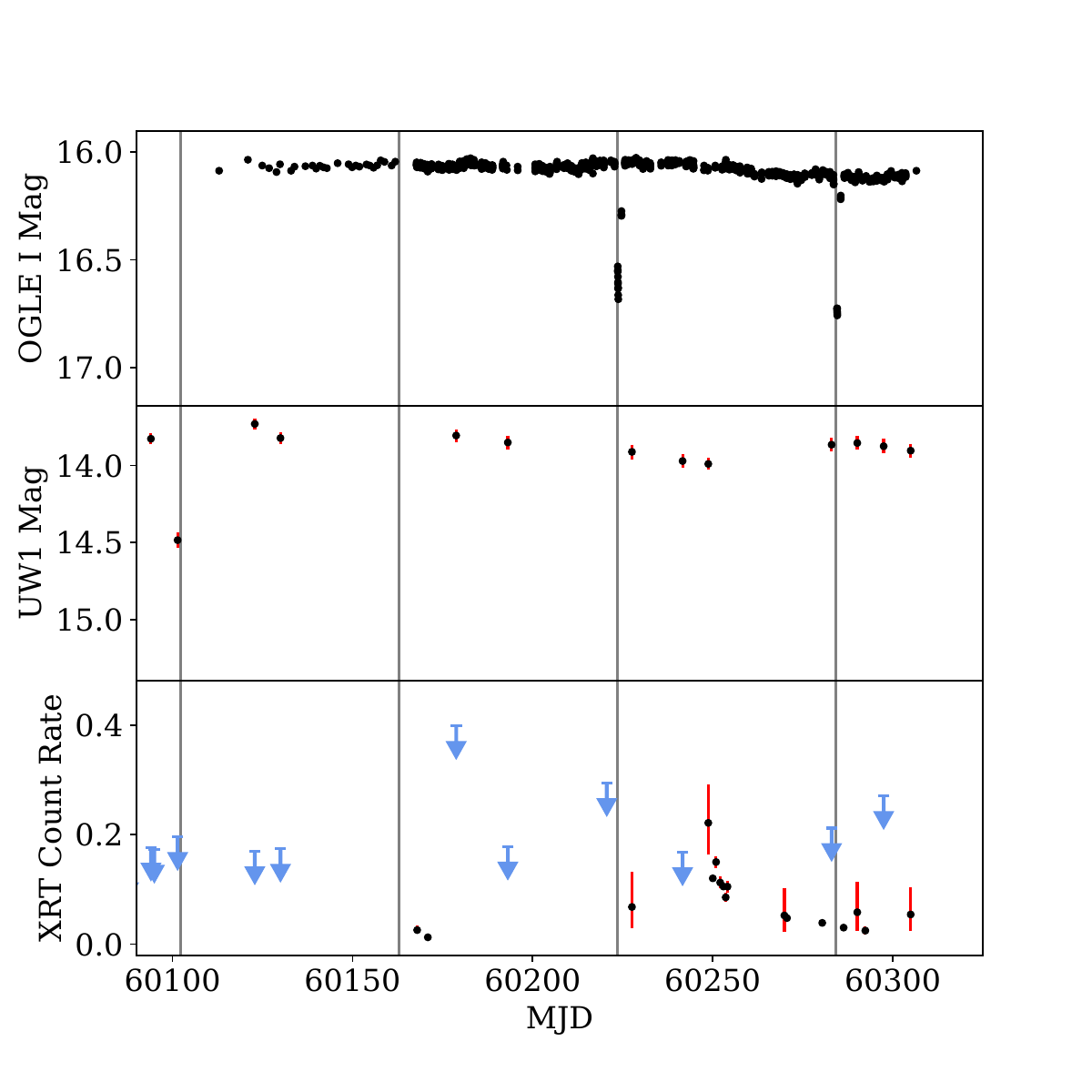}
    \caption{Truncated OGLE IV, Swift UVOT, and Swift XRT light curves for Swift J010902.6-723710 showing emission from MJD 60100 to MJD 60300. The times of optical eclipse, calculated using Equation \ref{eq:1}, are shown as vertical grey lines. Blue arrows show the calculated XRT flux upper limit during S-CUBED observations where Swift J010902.6-723710 is not detected.}
    \label{fig:trunc lc}
\end{figure}

The type of outburst observed in Swift J010902.6-723710 has remained uncertain as long-term monitoring of the source continues to produce XRT detections well past the peak of the initial outburst event. In Type I outbursts, X-ray emission is often limited both by the duration of an orbit and the current orbital phase. Emission is expected to be detected at a narrow range of values near the periastron passage of the orbit and is expected to last for no longer than a full orbital period \citep{1986Stella}. Additionally, Type I outbursts typically demonstrate a moderate increase in luminosity, peaking at $ L_X \sim 10^{36}-10^{37}$ ergs s$^{-1}$ \citep{2001Okazaki, 2011Reig}. Table \ref{tab:XRT data} shows the date of XRT detections of Swift J010902.6-723710 and the orbital phase at which they occurred with respect to Equation \ref{eq:1}. The light curve of XRT detections during outburst is plotted with respect to the time of optical eclipses in Figure \ref{fig:trunc lc}.  Both the figure and table show the unusually long duration of this outburst. The first S-CUBED detection of the outburst on October 10th occurred at Phase 0.06, which is approximately the periastron passage of orbit. However, other subsequent detections continue to occur well past the periastron passage of the orbit, demonstrating phase independence of the emission. The last two S-CUBED detections, occurring on December 12th and December 26th, correspond to Phase 0.13 and 0.36 of a new orbit. These subsequent detections are not consistent with the typical behavior demonstrated by a Type I outburst and have more in common with the emission signature of a Type II outburst \citep{2001Okazaki, 2017Townsend, 2022Tamang}. If this is indeed a Type II outburst, then the peak luminosity is much smaller than is typical for these outbursts. The peak X-ray luminosity of $L_X = 1.35 \times 10^{37}$ erg s$^{-1}$ is the only detection to occur above $10^{37}$ erg s$^{-1}$. Based on these features, one of two situations has occurred. One possibility is that the system has produced an abnormally long-duration Type I outburst with a typical peak luminosity \citep{2001Okazaki, 2011Reig}. Alternatively, the system has produced a Type II outburst that fails to reach the characteristic large peak luminosity that is expected during these events \citep{2001Okazaki, 2011Reig}.

However, it is important  to remember that this traditional classification scheme is very quantised - either it is called Type I or Type II. Since originally proposed in 1986 \citep{1986Stella} there have been many more diverse examples of X-ray outbursts from BeXRB systems observed. In reality there  must be a scope for a whole spectrum of outburst types since the outburst duration and the X-ray luminosity seen depend almost entirely on the interaction between the circumstellar disc and the neutron star. That in turn depends upon the characteristics of the circumstellar disc (density, size, inclination, mass outflow rate etc.) and the characteristics of the neutron star orbit (eccentricity, inclination to the circumstellar disk plane, phase of periastron etc.). So, given all those free parameters one should expect a whole range of observational characteristics, and not be driven to simply call them Type I or Type II.  SXP 182 is an excellent example of how the original classification scheme can be overly simplistic.

\section{Conclusion} \label{sec:conclusion}

This paper reports the detection of a previously-unknown BeXRB via weekly observations of the S-CUBED survey. This new system, Swift J010902.6-723710, was identified via a transient X-ray outburst and followed up via multi-wavelength observations. Deep follow-up X-ray observations identify a proposed spin period of 182s for the NS in this binary system. Historical light curve analysis of both UV and IR emission reveal the presence of strong eclipse-like features that re-appear every 60.623 days in the light curve, which is adopted as the proposed orbital period of the system. Optical spectroscopy reveals a strongly double-peaked H$\alpha$ emission line, indicating a highly-inclined system with an inclination of $i = 72-90^{\circ}$. Spectroscopic observations are also used to constrain the spectral class of the optical companion as a B0-0.5 star of spectral class Ve.

The proposed spin and orbital periods place Swift J010902.6-723710 place in the center of the expected distribution for similar BeXRBs on the Corbet Diagram. Eclipsing behavior is found to be caused by a $3.3 R_\odot$ accretion disk that surrounds the NS, making this the third eclipsing BeXRB to be detected so far. Finally, the type of outburst observed is found to be uncertain, with characteristics of both Type I and Type II outbursts being found in the X-ray emission of the source. More observations of this system are needed, particularly during subsequent eclipses, in order to further constrain the physical parameters of this rare eclipsing system and better understand this long-lasting, moderately luminous X-ray outburst. 

\section{acknowledgments}

This work made use of data supplied by the UK Swift Science Data Centre at the University of Leicester. JAK and TMG acknowledge the support of NASA contract NAS5-00136. We acknowledge the use of public data from the Swift data archive.

The SALT observations reported in this paper were obtained under the SALT Large Science Programme on transients (2021-2-LSP-001; PI: DAHB), which is also supported by Poland under grant no. MNiSW DIR/WK/2016/07.

PAE acknowledges UKSA support. DAHB acknowledges research support from the National Research Foundation. LJT is supported by the SALT Foundation and the NRF.

%

\vspace{5mm}
\facilities{Swift (XRT and UVOT), OGLE, SALT}





\bibliography{SC1772}{}

\begin{thebibliography}{}
\expandafter\ifx\csname natexlab\endcsname\relax\def\natexlab#1{#1}\fi
\providecommand{\url}[1]{\href{#1}{#1}}
\providecommand{\dodoi}[1]{doi:~\href{http://doi.org/#1}{\nolinkurl{#1}}}
\providecommand{\doeprint}[1]{\href{http://ascl.net/#1}{\nolinkurl{http://ascl.net/#1}}}
\providecommand{\doarXiv}[1]{\href{https://arxiv.org/abs/#1}{\nolinkurl{https://arxiv.org/abs/#1}}}

\bibitem[{{Allen}(1976)}]{1976allen}
{Allen}, C.~W. 1976, {Astrophysical Quantities}

\bibitem[{{Antoniou} {et~al.}(2010){Antoniou}, {Zezas}, {Hatzidimitriou}, \& {Kalogera}}]{2010Antoniou}
{Antoniou}, V., {Zezas}, A., {Hatzidimitriou}, D., \& {Kalogera}, V. 2010, \apjl, 716, L140, \dodoi{10.1088/2041-8205/716/2/L140}

\bibitem[{{Blackburn} {et~al.}(1999){Blackburn}, {Shaw}, {Payne}, {Hayes}, \& {Heasarc}}]{1999Blackburn}
{Blackburn}, J.~K., {Shaw}, R.~A., {Payne}, H.~E., {Hayes}, J.~J.~E., \& {Heasarc}. 1999, {FTOOLS: A general package of software to manipulate FITS files}, Astrophysics Source Code Library, record ascl:9912.002.
\newblock \doeprint{9912.002}

\bibitem[{{Buccheri} {et~al.}(1983){Buccheri}, {Bennett}, {Bignami}, {Bloemen}, {Boriakoff}, {Caraveo}, {Hermsen}, {Kanbach}, {Manchester}, {Masnou}, {Mayer-Hasselwander}, {{\"O}zel}, {Paul}, {Sacco}, {Scarsi}, \& {Strong}}]{1983Buccheri}
{Buccheri}, R., {Bennett}, K., {Bignami}, G.~F., {et~al.} 1983, \aap, 128, 245

\bibitem[{{Buckley} {et~al.}(2006){Buckley}, {Swart}, \& {Meiring}}]{2006Buckley}
{Buckley}, D.~A.~H., {Swart}, G.~P., \& {Meiring}, J.~G. 2006, in \procspie, Vol. 6267, Society of Photo-Optical Instrumentation Engineers (SPIE) Conference Series, 62670Z, \dodoi{10.1117/12.673750}

\bibitem[{{Burgh} {et~al.}(2003){Burgh}, {Nordsieck}, {Kobulnicky}, {Williams}, {O'Donoghue}, {Smith}, \& {Percival}}]{2003Burgh}
{Burgh}, E.~B., {Nordsieck}, K.~H., {Kobulnicky}, H.~A., {et~al.} 2003, in Society of Photo-Optical Instrumentation Engineers (SPIE) Conference Series, Vol. 4841, \procspie, ed. M.~{Iye} \& A.~F.~M. {Moorwood}, 1463--1471, \dodoi{10.1117/12.460312}

\bibitem[{{Burrows} {et~al.}(2005){Burrows}, {Hill}, {Nousek}, {Kennea}, {Wells}, {Osborne}, {Abbey}, {Beardmore}, {Mukerjee}, {Short}, {Chincarini}, {Campana}, {Citterio}, {Moretti}, {Pagani}, {Tagliaferri}, {Giommi}, {Capalbi}, {Tamburelli}, {Angelini}, {Cusumano}, {Br{\"a}uninger}, {Burkert}, \& {Hartner}}]{2005Burrows}
{Burrows}, D.~N., {Hill}, J.~E., {Nousek}, J.~A., {et~al.} 2005, \ssr, 120, 165, \dodoi{10.1007/s11214-005-5097-2}

\bibitem[{{Carpano} {et~al.}(2022){Carpano}, {Haberl}, {Maitra}, {Freyberg}, {Dennerl}, {Schwope}, {Buckley}, \& {Monageng}}]{2022Carpano}
{Carpano}, S., {Haberl}, F., {Maitra}, C., {et~al.} 2022, \aap, 661, A20, \dodoi{10.1051/0004-6361/202141082}

\bibitem[{{Carpano} {et~al.}(2017){Carpano}, {Haberl}, \& {Sturm}}]{2017Carpano}
{Carpano}, S., {Haberl}, F., \& {Sturm}, R. 2017, \aap, 602, A81, \dodoi{10.1051/0004-6361/201629299}

\bibitem[{{Coe} {et~al.}(2015){Coe}, {Bartlett}, {Bird}, {Haberl}, {Kennea}, {McBride}, {Townsend}, \& {Udalski}}]{2015Coe2}
{Coe}, M.~J., {Bartlett}, E.~S., {Bird}, A.~J., {et~al.} 2015, \mnras, 447, 2387, \dodoi{10.1093/mnras/stu2568}

\bibitem[{{Coe} {et~al.}(2021){Coe}, {Kennea}, {Evans}, {Townsend}, {Udalski}, {Monageng}, \& {Buckley}}]{2021Coe}
{Coe}, M.~J., {Kennea}, J.~A., {Evans}, P.~A., {et~al.} 2021, \mnras, 504, 1398, \dodoi{10.1093/mnras/stab972}

\bibitem[{{Coe} {et~al.}(2023{\natexlab{a}}){Coe}, {Kennea}, {Gaudin}, {Monageng}, {Townsend}, {Buckley}, {Udalski}, \& {Evans}}]{2023bcoe}
{Coe}, M.~J., {Kennea}, J.~A., {Gaudin}, T.~M., {et~al.} 2023{\natexlab{a}}, The Astronomer's Telegram, 16321, 1

\bibitem[{{Coe} {et~al.}(2023{\natexlab{b}}){Coe}, {Kennea}, {Monageng}, {Buckley}, {Udalski}, \& {Evans}}]{2023Coe}
{Coe}, M.~J., {Kennea}, J.~A., {Monageng}, I.~M., {et~al.} 2023{\natexlab{b}}, \mnras, 524, 3263, \dodoi{10.1093/mnras/stad1987}

\bibitem[{{Coe} \& {Kirk}(2015)}]{2015coekirk}
{Coe}, M.~J., \& {Kirk}, J. 2015, \mnras, 452, 969, \dodoi{10.1093/mnras/stv1283}

\bibitem[{{Corbet}(1984)}]{1984Corbet}
{Corbet}, R.~H.~D. 1984, \aap, 141, 91

\bibitem[{{Cox}(2000)}]{Cox2000}
{Cox}, A.~N. 2000, {Allen's astrophysical quantities}

\bibitem[{{Crawford} {et~al.}(2012){Crawford}, {Still}, {Schellart}, {Balona}, {Buckley}, {Gulbis}, {Kniazev}, {Kotze}, {Loaring}, {Nordsieck}, {Pickering}, {Potter}, {Romero Colmenero}, {Vaisanen}, {Wiliams}, \& {Zietsman}}]{2012Crawford}
{Crawford}, S.~M., {Still}, M., {Schellart}, P., {et~al.} 2012, {PySALT: SALT science pipeline}, Astrophysics Source Code Library.
\newblock \doeprint{1207.010}

\bibitem[{{Evans} {et~al.}(2004){Evans}, {Howarth}, {Irwin}, {Burnley}, \& {Harries}}]{Evans2004}
{Evans}, C.~J., {Howarth}, I.~D., {Irwin}, M.~J., {Burnley}, A.~W., \& {Harries}, T.~J. 2004, \mnras, 353, 601, \dodoi{10.1111/j.1365-2966.2004.08096.x}

\bibitem[{{Evans} {et~al.}(2014){Evans}, {Osborne}, {Beardmore}, {Page}, {Willingale}, {Mountford}, {Pagani}, {Burrows}, {Kennea}, {Perri}, {Tagliaferri}, \& {Gehrels}}]{2014Evans}
{Evans}, P.~A., {Osborne}, J.~P., {Beardmore}, A.~P., {et~al.} 2014, \apjs, 210, 8, \dodoi{10.1088/0067-0049/210/1/8}

\bibitem[{{Gehrels} {et~al.}(2004){Gehrels}, {Chincarini}, {Giommi}, {Mason}, {Nousek}, {Wells}, {White}, {Barthelmy}, {Burrows}, {Cominsky}, {Hurley}, {Marshall}, {M{\'e}sz{\'a}ros}, {Roming}, {Angelini}, {Barbier}, {Belloni}, {Campana}, {Caraveo}, {Chester}, {Citterio}, {Cline}, {Cropper}, {Cummings}, {Dean}, {Feigelson}, {Fenimore}, {Frail}, {Fruchter}, {Garmire}, {Gendreau}, {Ghisellini}, {Greiner}, {Hill}, {Hunsberger}, {Krimm}, {Kulkarni}, {Kumar}, {Lebrun}, {Lloyd-Ronning}, {Markwardt}, {Mattson}, {Mushotzky}, {Norris}, {Osborne}, {Paczynski}, {Palmer}, {Park}, {Parsons}, {Paul}, {Rees}, {Reynolds}, {Rhoads}, {Sasseen}, {Schaefer}, {Short}, {Smale}, {Smith}, {Stella}, {Tagliaferri}, {Takahashi}, {Tashiro}, {Townsley}, {Tueller}, {Turner}, {Vietri}, {Voges}, {Ward}, {Willingale}, {Zerbi}, \& {Zhang}}]{2004Gehrels}
{Gehrels}, N., {Chincarini}, G., {Giommi}, P., {et~al.} 2004, \apj, 611, 1005, \dodoi{10.1086/422091}

\bibitem[{{Graczyk} {et~al.}(2013){Graczyk}, {Pietrzy{\'n}ski}, {Pilecki}, {Thompson}, {Gieren}, {Konorski}, {Udalski}, \& {Soszy{\'n}ski}}]{Graczyk2013}
{Graczyk}, D., {Pietrzy{\'n}ski}, G., {Pilecki}, B., {et~al.} 2013, in IAU Symposium, Vol. 289, Advancing the Physics of Cosmic Distances, ed. R.~{de Grijs}, 222--225, \dodoi{10.1017/S1743921312021436}

\bibitem[{{Graczyk} {et~al.}(2020){Graczyk}, {Pietrzy{\'n}ski}, {Thompson}, {Gieren}, {Zgirski}, {Villanova}, {G{\'o}rski}, {Wielg{\'o}rski}, {Karczmarek}, {Narloch}, {Pilecki}, {Taormina}, {Smolec}, {Suchomska}, {Gallenne}, {Nardetto}, {Storm}, {Kudritzki}, {Ka{\l}uszy{\'n}ski}, \& {Pych}}]{2020Graczyk}
{Graczyk}, D., {Pietrzy{\'n}ski}, G., {Thompson}, I.~B., {et~al.} 2020, \apj, 904, 13, \dodoi{10.3847/1538-4357/abbb2b}

\bibitem[{{Haberl} \& {Pietsch}(2005)}]{2005Haberl}
{Haberl}, F., \& {Pietsch}, W. 2005, \aap, 438, 211, \dodoi{10.1051/0004-6361:20042470}

\bibitem[{{Haberl} \& {Sturm}(2016)}]{2016Haberl}
{Haberl}, F., \& {Sturm}, R. 2016, \aap, 586, A81, \dodoi{10.1051/0004-6361/201527326}

\bibitem[{{Harmanec}(1983)}]{Harmanec1983}
{Harmanec}, P. 1983, Hvar Observatory Bulletin, 7, 55

\bibitem[{{Harris} \& {Zaritsky}(2004)}]{2004Harris}
{Harris}, J., \& {Zaritsky}, D. 2004, \aj, 127, 1531, \dodoi{10.1086/381953}

\bibitem[{{Huang}(1972)}]{Huang1972}
{Huang}, S.-S. 1972, \apj, 171, 549, \dodoi{10.1086/151309}

\bibitem[{{Jaisawal} {et~al.}(2023){Jaisawal}, {Vasilopoulos}, {Naik}, {Maitra}, {Malacaria}, {Chhotaray}, {Gendreau}, {Guillot}, {Ng}, \& {Sanna}}]{2023Jaisawal}
{Jaisawal}, G.~K., {Vasilopoulos}, G., {Naik}, S., {et~al.} 2023, \mnras, 521, 3951, \dodoi{10.1093/mnras/stad781}

\bibitem[{{Kennea} {et~al.}(2020){Kennea}, {Coe}, {Evans}, {Monageng}, {Townsend}, {Siegel}, {Udalski}, \& {Buckley}}]{2020Kennea}
{Kennea}, J.~A., {Coe}, M.~J., {Evans}, P.~A., {et~al.} 2020, \mnras, 499, L41, \dodoi{10.1093/mnrasl/slaa154}

\bibitem[{{Kennea} {et~al.}(2021){Kennea}, {Coe}, {Evans}, {Townsend}, {Campbell}, \& {Udalski}}]{2021Kennea}
---. 2021, \mnras, 508, 781, \dodoi{10.1093/mnras/stab2632}

\bibitem[{{Kennea} {et~al.}(2018){Kennea}, {Coe}, {Evans}, {Waters}, \& {Jasko}}]{2018Kennea}
{Kennea}, J.~A., {Coe}, M.~J., {Evans}, P.~A., {Waters}, J., \& {Jasko}, R.~E. 2018, \apj, 868, 47, \dodoi{10.3847/1538-4357/aae839}

\bibitem[{{Kobulnicky} {et~al.}(2003){Kobulnicky}, {Nordsieck}, {Burgh}, {Smith}, {Percival}, {Williams}, \& {O'Donoghue}}]{2003Kobulnicky}
{Kobulnicky}, H.~A., {Nordsieck}, K.~H., {Burgh}, E.~B., {et~al.} 2003, in \procspie, Vol. 4841, Instrument Design and Performance for Optical/Infrared Ground-based Telescopes, ed. M.~{Iye} \& A.~F.~M. {Moorwood}, 1634--1644, \dodoi{10.1117/12.460315}

\bibitem[{{Lazzarini} {et~al.}(2019){Lazzarini}, {Williams}, {Hornschemeier}, {Antoniou}, {Vasilopoulos}, {Haberl}, {Vulic}, {Yukita}, {Zezas}, {Bodaghee}, {Lehmer}, {Maccarone}, {Ptak}, {Wik}, {Fornasini}, {Hong}, {Kennea}, {Tomsick}, {Venters}, {Udalski}, \& {Cassity}}]{2019Lazzarini}
{Lazzarini}, M., {Williams}, B.~F., {Hornschemeier}, A.~E., {et~al.} 2019, \apj, 884, 2, \dodoi{10.3847/1538-4357/ab3f32}

\bibitem[{{Maggi} {et~al.}(2013){Maggi}, {Haberl}, {Sturm}, {Pietsch}, {Rau}, {Greiner}, {Udalski}, \& {Sasaki}}]{2013Maggi}
{Maggi}, P., {Haberl}, F., {Sturm}, R., {et~al.} 2013, \aap, 554, A1, \dodoi{10.1051/0004-6361/201321238}

\bibitem[{{Maitra} {et~al.}(2023){Maitra}, {Haberl}, {Kaltenbrunner}, {Doroshenko}, {Ducci}, \& {Udalski}}]{2023Maitra}
{Maitra}, C., {Haberl}, F., {Kaltenbrunner}, D., {et~al.} 2023, The Astronomer's Telegram, 15886, 1

\bibitem[{{Martin} {et~al.}(2014){Martin}, {Nixon}, {Armitage}, {Lubow}, \& {Price}}]{2014Martin}
{Martin}, R.~G., {Nixon}, C., {Armitage}, P.~J., {Lubow}, S.~H., \& {Price}, D.~J. 2014, \apjl, 790, L34, \dodoi{10.1088/2041-8205/790/2/L34}

\bibitem[{{McBride} {et~al.}(2017){McBride}, {Gonz{\'a}lez-Gal{\'a}n}, {Bird}, {Coe}, {Bartlett}, {Dorda}, {Haberl}, {Marco}, {Negueruela}, {Schurch}, {Sturm}, {Buckley}, \& {Udalski}}]{2017McBride}
{McBride}, V.~A., {Gonz{\'a}lez-Gal{\'a}n}, A., {Bird}, A.~J., {et~al.} 2017, \mnras, 467, 1526, \dodoi{10.1093/mnras/stx181}

\bibitem[{{McConnachie}(2012)}]{McConnachie2012}
{McConnachie}, A.~W. 2012, \aj, 144, 4, \dodoi{10.1088/0004-6256/144/1/4}

\bibitem[{{Monageng} {et~al.}(2017){Monageng}, {McBride}, {Coe}, {Steele}, \& {Reig}}]{2017Monageng}
{Monageng}, I.~M., {McBride}, V.~A., {Coe}, M.~J., {Steele}, I.~A., \& {Reig}, P. 2017, \mnras, 464, 572, \dodoi{10.1093/mnras/stw2354}

\bibitem[{{Monageng} {et~al.}(2020){Monageng}, {Coe}, {Buckley}, {McBride}, {Kennea}, {Udalski}, {Evans}, {Clark}, \& {Negueruela}}]{2020Monageng}
{Monageng}, I.~M., {Coe}, M.~J., {Buckley}, D.~A.~H., {et~al.} 2020, \mnras, 496, 3615, \dodoi{10.1093/mnras/staa1739}

\bibitem[{{Monageng} {et~al.}(2022){Monageng}, {Coe}, {Townsend}, {Laycock}, {Kennea}, {Roy}, {Udalski}, {Bhattacharya}, {Christodoulou}, {Buckley}, \& {Evans}}]{2022Monageng}
{Monageng}, I.~M., {Coe}, M.~J., {Townsend}, L.~J., {et~al.} 2022, \mnras, 511, 6075, \dodoi{10.1093/mnras/stac106}

\bibitem[{{Okazaki} \& {Negueruela}(2001)}]{2001Okazaki}
{Okazaki}, A.~T., \& {Negueruela}, I. 2001, \aap, 377, 161, \dodoi{10.1051/0004-6361:20011083}

\bibitem[{{Pecaut} \& {Mamajek}(2013)}]{Pecaut2013}
{Pecaut}, M.~J., \& {Mamajek}, E.~E. 2013, \apjs, 208, 9, \dodoi{10.1088/0067-0049/208/1/9}

\bibitem[{{Rajoelimanana} {et~al.}(2011){Rajoelimanana}, {Charles}, \& {Udalski}}]{Rajoelimanana2011}
{Rajoelimanana}, A.~F., {Charles}, P.~A., \& {Udalski}, A. 2011, \mnras, 413, 1600, \dodoi{10.1111/j.1365-2966.2011.18243.x}

\bibitem[{{Reig}(2011)}]{2011Reig}
{Reig}, P. 2011, \apss, 332, 1, \dodoi{10.1007/s10509-010-0575-8}

\bibitem[{{Reig} \& {Fabregat}(2015)}]{Reig2015}
{Reig}, P., \& {Fabregat}, J. 2015, \aap, 574, A33, \dodoi{10.1051/0004-6361/201425008}

\bibitem[{{Rezaeikh} {et~al.}(2014){Rezaeikh}, {Javadi}, {Khosroshahi}, \& {van Loon}}]{2014Rezaeikh}
{Rezaeikh}, S., {Javadi}, A., {Khosroshahi}, H., \& {van Loon}, J.~T. 2014, \mnras, 445, 2214, \dodoi{10.1093/mnras/stu1807}

\bibitem[{{Rivinius}(2019)}]{2019Rivinius}
{Rivinius}, T. 2019, IAU Symposium, 346, 105, \dodoi{10.1017/S1743921318008207}

\bibitem[{{Roming} {et~al.}(2005){Roming}, {Kennedy}, {Mason}, {Nousek}, {Ahr}, {Bingham}, {Broos}, {Carter}, {Hancock}, {Huckle}, {Hunsberger}, {Kawakami}, {Killough}, {Koch}, {McLelland}, {Smith}, {Smith}, {Soto}, {Boyd}, {Breeveld}, {Holland}, {Ivanushkina}, {Pryzby}, {Still}, \& {Stock}}]{2005Roming}
{Roming}, P. W.~A., {Kennedy}, T.~E., {Mason}, K.~O., {et~al.} 2005, \ssr, 120, 95, \dodoi{10.1007/s11214-005-5095-4}

\bibitem[{{Roy} {et~al.}(2022){Roy}, {Cappallo}, {Laycock}, {Christodoulou}, {Vasilopoulos}, \& {Bhattacharya}}]{2022Roy}
{Roy}, A., {Cappallo}, R., {Laycock}, S. G.~T., {et~al.} 2022, \apj, 936, 90, \dodoi{10.3847/1538-4357/ac82b6}

\bibitem[{{Stella} {et~al.}(1986){Stella}, {White}, \& {Rosner}}]{1986Stella}
{Stella}, L., {White}, N.~E., \& {Rosner}, R. 1986, \apj, 308, 669, \dodoi{10.1086/164538}

\bibitem[{{Straizys} \& {Kuriliene}(1981)}]{Straizys1981}
{Straizys}, V., \& {Kuriliene}, G. 1981, \apss, 80, 353, \dodoi{10.1007/BF00652936}

\bibitem[{{Tamang} {et~al.}(2022){Tamang}, {Ghising}, {Tobrej}, {Rai}, \& {Paul}}]{2022Tamang}
{Tamang}, R., {Ghising}, M., {Tobrej}, M., {Rai}, B., \& {Paul}, B.~C. 2022, \mnras, 515, 5407, \dodoi{10.1093/mnras/stac2135}

\bibitem[{{Townsend} {et~al.}(2011){Townsend}, {Coe}, {Corbet}, \& {Hill}}]{2011Townsend}
{Townsend}, L.~J., {Coe}, M.~J., {Corbet}, R.~H.~D., \& {Hill}, A.~B. 2011, \mnras, 416, 1556, \dodoi{10.1111/j.1365-2966.2011.19153.x}

\bibitem[{{Townsend} {et~al.}(2017){Townsend}, {Kennea}, {Coe}, {McBride}, {Buckley}, {Evans}, \& {Udalski}}]{2017Townsend}
{Townsend}, L.~J., {Kennea}, J.~A., {Coe}, M.~J., {et~al.} 2017, \mnras, 471, 3878, \dodoi{10.1093/mnras/stx1865}

\bibitem[{{Udalski} {et~al.}(2015){Udalski}, {Szyma{\'n}ski}, \& {Szyma{\'n}ski}}]{Udalski2015}
{Udalski}, A., {Szyma{\'n}ski}, M.~K., \& {Szyma{\'n}ski}, G. 2015, \actaa, 65, 1.
\newblock \doarXiv{1504.05966}

\bibitem[{{Willingale} {et~al.}(2013){Willingale}, {Starling}, {Beardmore}, {Tanvir}, \& {O'Brien}}]{2013Willignale}
{Willingale}, R., {Starling}, R.~L.~C., {Beardmore}, A.~P., {Tanvir}, N.~R., \& {O'Brien}, P.~T. 2013, \mnras, 431, 394, \dodoi{10.1093/mnras/stt175}

\end{thebibliography}
\bibliographystyle{aasjournal}



\end{document}